\documentclass[aps,prd,a4paper,nofootinbib,
preprintnumbers,superscriptaddress,showpacs,showkeys,twocolumn,10pt]{revtex4-1}

\usepackage{amsmath}
\usepackage{amssymb}
\usepackage{bm}
\usepackage{graphicx}
\usepackage{subcaption} 
\usepackage{color}
\usepackage[normalem]{ulem}  

\usepackage{empheq}

\newcommand{\be}{\begin{equation}}
\newcommand{\ee}{\end{equation}}
\newcommand{\ba}{\begin{eqnarray}}
\newcommand{\ea}{\end{eqnarray}}

\newcommand{\dd}{\mathrm{d}}
\newcommand{\mycomment}[1]{}

\renewcommand\sout{\bgroup \color{blue} \ULdepth=-.5ex \ULset}

\begin{document}

\title{Non-Markovian heavy-quark equilibration and \\ equilibrium correlation function in a thermal medium}

\author{Monisha Nair}
\affiliation{School of Physical Sciences, Indian Institute of Technology Goa, Ponda-403401, Goa, India}

\author{Juan M. Torres-Rincon}
\affiliation{Departament de F\'isica Qu\`antica i Astrof\'isica and Institut de Ci\`encies del Cosmos (ICCUB), Facultat de F\'isica,  Universitat de Barcelona, Mart\'i i Franqu\`es 1, Barcelona, 08028, Spain}

\author{Santosh K. Das}
\affiliation{School of Physical Sciences, Indian Institute of Technology Goa, Ponda-403401, Goa, India}


\begin{abstract}
We compute the charm-quark current-current correlation function within the Langevin framework and extract the heavy-quark diffusion coefficient using the Green--Kubo formula. The formalism is further extended to evaluate the correlation function using a generalized Langevin equation that incorporates memory effects via an exponentially decaying memory kernel. We find that while memory effects qualitatively modify the transient structure of the current correlations, the value of the transport coefficient remains unchanged. In addition, we investigate heavy-quark thermalization in the presence of memory and compare the nonequilibrium solution of a generalized Fokker--Planck equation with the one obtained from the generalized Langevin equation in the non-relativistic limit. We also consider the relativistic case with correlated noise and observe that memory effects can give rise to a damped, oscillatory equilibration of the heavy quark momentum distribution in a non-Markovian bath. 
\end{abstract}



\maketitle

\section{Introduction}

According to quantum chromodynamics (QCD), the fundamental theory of strong interactions, nuclear matter undergoes a phase transition at high temperature and density. Hadrons dissolve into an interacting medium of quarks and gluons, the quark-gluon plasma (QGP)~\cite{Shuryak:1980tp, Jacak:2012dx}. Experimental efforts are underway to create and characterize this phase of matter at the Relativistic Heavy Ion Collider and the Large Hadron Collider energies. Probing the QGP remains a topic of significant interest. The dynamics of the QGP are governed primarily by light quarks and gluons, along with a smaller population of heavy quarks. Heavy quarks (HQs)~\cite{Das:2024vac,He:2022ywp, Rapp:2018qla, Dong:2019unq, Cao:2018ews}, mainly charm and beauty quarks, are regarded as ideal probes of the QGP formed in high-energy nuclear collisions. Owing to their large masses, $m$, they are produced early in the initial hard scatterings, with a formation time 
$\tau_{\mathrm{formation}} \sim 1/(2m)$,
allowing them to experience the entire space--time evolution of the system.
Furthermore, the thermalization time of heavy quarks is delayed relative to that of light partons in the bulk medium by a factor of the order of $m/T$, making their thermalization timescale comparable to the lifetime of the QGP fireball. As a result, heavy quarks are not expected to fully thermalize and therefore retain information about their interaction history. The propagation of HQs in the QGP can be described as an interaction between equilibrium and nonequilibrium degrees of freedom. In the thermal medium of light quarks and gluons, heavy quarks effectively  undergo Brownian motion, and their evolution can be studied using Langevin or Fokker--Planck formalisms~\cite{Svetitsky:1987gq,Moore:2004tg,vanHees:2007me}.

The standard approach to studying HQ dynamics in the QGP is to track the evolution of their position and momentum using the Langevin equation within an expanding thermal medium. The HQ drag and diffusion coefficients in the QGP, which encode the detailed interactions of heavy quarks with the thermal medium, serve as input to the Langevin equation. Solving the Langevin equation allows one to compute various experimental observables, such as the nuclear suppression factor \(R_{AA}\) and the elliptic flow \(v_2\)~\cite{vanHees:2007me,Lang:2012nqy,Das:2015ana,Cao:2015hia,Beraudo:2014boa,Singh:2025duj}. The values of the transport coefficients that best reproduce the experimental observables can then be used to extract the spatial diffusion coefficient $D_s$~\cite{He:2012df,Ozvenchuk:2014rpa,Das:2016llg,Scardina:2017ipo,Grishmanovskii:2025mnc, ALICE:2021rxa,Sambataro:2025pop} in the $p \to 0$ limit, which can be directly compared with lattice-QCD results. Thus, HQs act as powerful probes that connect phenomenological modeling with experimental measurements and, ultimately, with lattice-QCD results. 

In Langevin dynamics, a key issue is how to model thermal noise $\eta$. 
This noise is stochastic in nature and its strength is regulated by the diffusion coefficient in momentum space, $B$. It is typically modeled as a standard Wiener process, which implies the absence of temporal correlations, with $\langle \eta(t) \rangle = 0$ and 
$\langle \eta(t)\eta(t') \rangle = 2B\,\delta(t - t')$. Such noise is referred to as white noise. It represents a memoryless fluctuating field in which successive thermal kicks are uncorrelated in time. In other words, their correlations decay instantaneously, as expressed by the Dirac delta-function  correlation function. Most studies of heavy quarks in the QGP are carried out within the white-noise approximation~\cite{vanHees:2007me,Lang:2012nqy, Das:2015ana,Cao:2015hia, Beraudo:2014boa, Singh:2025duj,He:2012df,Ozvenchuk:2014rpa,Das:2016llg}.

However, several recent studies have gone beyond this approximation and considered scenarios in which thermal kicks are time-correlated~\cite{Schenke:2006uh,Kapusta:2011gt,Kapusta:2012zb,Schmidt:2014zpa,Murase:2016rhl,Kapusta:2017hfi,Hammelmann:2018ath,Schuller:2019ega,Chen:2023pgx,Oliveira:2019kkk, Ruggieri:2019zos,Ruggieri:2022kxv, Pooja:2023gqt,Prakash:2024rdz,Prakash:2024irm,Prakash:2026izm,Metzler:2000prr}. Such stochastic processes are referred to as processes with memory. In this context, hydrodynamic fluctuations~\cite{Kapusta:2011gt,Murase:2016rhl}, heavy quark diffusion in the QGP~\cite{Ruggieri:2022kxv,Pooja:2023gqt,Prakash:2024rdz,Prakash:2026izm} and the evolving Glasma~\cite{Boguslavski:2020tqz,Liu:2019lac,Liu:2020cpj,Khowal:2021zoo,Avramescu:2023qvv}, electric charge diffusion~\cite{Kapusta:2017hfi}, dilepton production~\cite{Schenke:2006uh}, and the electrical conductivity of the hadron gas~\cite{Hammelmann:2018ath} are among the physical problems in which memory effects may play an important role.

Studies accounting for memory in heavy-quark dynamics suggest that non-Markovian effects may significantly influence heavy-flavor observables such as $R_{AA}$~\cite{Ruggieri:2022kxv}. However, most recent studies focus primarily on HQ momentum evolution in the presence of memory. In this paper, we investigate HQ dynamics in a thermal bath while explicitly incorporating memory effects. We assume a specific form of noise correlation that decays exponentially with time. The main focus is to study heavy-quark thermalization and the current--current correlation function in order to extract the HQ spatial diffusion coefficient, $D_s$, using the Green--Kubo relation in the presence of memory, thereby assessing the impact of non-Markovian HQ dynamics in a static QGP medium. A generalized Fokker--Planck equation with memory is also considered to compare with the results obtained from the generalized Langevin equation with memory and to demonstrate the approach to thermalization in the presence of memory.

In a non-relativistic context, non-Markovian dynamics has been extensively studied for more than 50 years~\cite{adelman1976fokker,fox1977generalized,fox1978gaussian}. In this work, we present some analytical results in this domain. However, we primarily consider the relativistic dynamics in the numerical application of the Langevin equation, where analytical expressions  are generally unavailable~\cite{Dunkel:2008ngc}. The numerical scheme to implement the exponential memory kernel is adopted from a previous reference by one of us~\cite{Ruggieri:2022kxv}. However, in this case, we apply an energy-dependent diffusion coefficient $B(E)$, for which the noise naturally turns out to be multiplicative, increasing the difficulty in solving the Langevin system. This issue has been treated in Ref.~\cite{He:2013zua}, but only in the standard white-noise formalism. In this paper, we unify non-Markovian and relativistic dynamics, both in equilibrium and out of equilibrium. In doing this, we describe how the Wong-Zakai theorem~\cite{wong1969riemann} forces us to adopt a particular numerical scheme for the generalized Langevin equation. To present a complete picture, we also investigate the Markovian and non-relativistic limits in both equilibrium and nonequilibrium, thus providing a solid framework for future works.

\section{The Langevin equation with Gaussian white noise}
\label{LV_wn}
 
In this section, we describe the theoretical formalism of microscopic motion of HQs, the Langevin equation, and its mesoscopic counterpart in terms of the Fokker--Planck equation. We also consider a macroscopic description in terms of a heavy-flavor current, obtained by coarse-graining the microscopic motion of HQs. Using the theory of hydrodynamic fluctuations~\cite{Landau,Kapusta:2011gt} we can derive the Green--Kubo formula, which captures the diffusion coefficient in terms of the current--current correlation function in equilibrium. Details of this formalism are given in App.~\ref{app:hydrofluc}, applied to a single conserved (heavy-flavor) current.

\subsection{Langevin equation without memory}

To describe HQ dynamics within a thermal bath, such as the QGP, we consider the Langevin equation~\cite{van1983stochastic}. It is a stochastic differential equation that models the Brownian-like motion of heavy quarks in a thermalized medium. It reads,
\begin{empheq}[left=\empheqlbrace]{align}
\frac{\dd x^i}{\dd t} &= \frac{p^i}{E} \label{eq:langevinbis} \\
\frac{\dd p^i }{\dd t} & = -\Gamma(p) p^i  + \eta^i (t) \ , \label{eq:langevin}
\end{empheq}
where $\bm{p}$ is the 3-momentum of the HQ (the index $i$ represents the Cartesian coordinate), $E=\sqrt{p^2+m^2}$, $m$ the heavy-particle mass, and $\Gamma(p)$ is the friction coefficient; which, in general, is a momentum (and temperature)-dependent quantity~\cite{He:2013zua}. Finally, the noise $\eta(t)$ is a stochastic term, which represents random momentum kicks from the constituents of the medium (microscopically, light quarks and gluons).

Within this section, we consider the noise to be Gaussian (with a distribution entirely defined by its first two moments) and ``white'' or uncorrelated. The latter approximation simplifies the analysis significantly, as the noise correlations become a Dirac delta function,
\begin{empheq}[left=\empheqlbrace]{align}
\langle \eta^i(t) \rangle &= 0 \ , \\
\langle \eta^i(t) \eta^j(t') \rangle &= 2B(p) \delta^{ij} \delta(t - t') \ .
\label{eq:noise_corr}
\end{empheq}
This property leads to Markovian dynamics of the HQ. The coefficient $B(p)$ is the diffusion coefficient in momentum space. In writing Eq.~\eqref{eq:noise_corr}, we have implicitly assumed an isotropic and homogeneous medium, so that only one independent diffusion coefficient exists. For a more general description, we refer the reader to Ref.~\cite{He:2013zua}.

The white noise assumption implies that the HQ momentum fluctuations coming from medium interaction are instantaneous and memoryless, an idealization valid when the correlation time of the medium is much shorter than the characteristic timescale of the HQ's relaxation. In this paper, ``white'', ``memoryless'' and ``uncorrelated'' noise are treated as synonyms.

Since the coefficient $B(p)$ also depends in general on the HQ momentum, it is necessary to settle on the value of the momentum to be evaluated. The Langevin equation itself is not well defined until a prescription is taken. In terms of a real parameter $\zeta \in [0,1]$ the diffusion coefficient is to be evaluated as
\be B(\bm{p}+\zeta \dd\bm{p}) \ , \ee
where $\zeta=0$ is the so-called pre-point or Itô scheme; $\zeta=1/2$ is the mid-point or Stratonovich-Fisk choice; and $\zeta=1$ is the post-point or H\"anggi-Klimontovich prescription~\cite{dunkel2005theory}.

For white noise, each of these prescriptions is equally valid, as long as a consistent fluctuation--dissipation relation linking $\Gamma (p)$ and $B(p)$ is chosen accordingly~\cite{He:2013zua},
\be \Gamma(E)=\frac{1}{E} \left( \frac{B(E)}{T} - (1-\zeta) \frac{\partial B(E)}{\partial E} \right) \ , \label{eq:fricforce} \ee
with the momentum dependence hidden within the particle's energy $E=E(p)$.

In Ref.~\cite{He:2013zua}, the pre-point and post-point prescriptions were applied. In the former case, the evaluation of the Langevin step is simpler, since $B(E)$ is already fixed at the initial time step. In the latter case, the Langevin step is not direct, since one needs to know the value of the diffusion coefficient at the final momentum, which is not known \textit{a priori}. The same happens for the  mid-point rule (or Stratonovich-Fisk choice), $\zeta=1/2$, which is the one we use in this work (for reasons which will become clear later). In these cases, predictor-corrector numerical methods~\cite{toral2014stochastic} are typically used in practice~\cite{He:2013zua,Oei:2024pva}.

\subsection{The Fokker--Planck equation and equilibration}

The microscopic Langevin equation with white noise for the HQ momenta can be transformed into a Fokker--Planck (FP) equation for the HQ distribution function. It describes the time evolution of the probability distribution function $f(\bm{p},t)$ of the HQ momentum. This can be done by considering the Kramers--Moyal expansion of the master equation, which, under the assumption of small momentum transfers, can be truncated at second order.
It can also be derived from the Boltzmann equation by performing a similar expansion of the collision term~\cite{Das:2024vac}. As before, we assume spatial homogeneity so that the distribution function does not depend on the coordinates.

Under these assumptions, the FP equation in three dimensions is~\cite{toral2014stochastic,He:2013zua,Das:2024vac}
\begin{equation}
\frac{\partial f (\bm{p},t)}{\partial t} = \frac{\partial}{\partial p^i} \left[ A(E) p^i f(\bm{p},t) +  \frac{\partial [B(E) f(\bm{p},t)]}{\partial p^i} \right] \ ,
\label{eq:FP3D}
\end{equation}
where $A(E)$ is the drag coefficient and $B(E)$ is the momentum diffusion coefficient, already appearing in the Langevin equation~\cite{He:2013zua} (again, in the isotropic and homogeneous case).

The relation between $A(E)$ and $B(E)$ is established from the fluctuation-dissipation theorem, emerging from the condition that the stationary solution of the FP equation ($\partial f/\partial t = 0$) coincides with the equilibrium distribution function. The condition, also known as the Einstein relation, reads~\cite{Walton:1999dy,Das:2013kea}
\begin{equation}
    A(E) = \frac{1}{E} \left( \frac{B(E)}{T} - \frac{\partial B(E)}{\partial E} \right) \ . \label{eq:Einsteingeneral}
\end{equation}

In the non-relativistic regime, one can make both $A$ and $B$ independent of the energy and set,
\begin{equation}
B = A m T \quad (\text{non-relativistic}) \ . \label{eq:EinsteinNR} 
\end{equation}
In the relativistic domain, at least one of the coefficients must be energy-dependent~\cite{He:2013zua}. For simplicity, one could choose a constant $B$, so that the noise in the Langevin equation~\eqref{eq:langevinbis}-\eqref{eq:langevin} remains additive, instead of multiplicative. However, for convenience in this work, we set a constant $A$, whose inverse, a relaxation time, fixes the fundamental time scale in the system, independent of $E$ and $T$. With this choice, to satisfy Eq.~\eqref{eq:Einsteingeneral} in the relativistic case, we use the particular form
\begin{equation}
    B(E) = A (E+T) T \quad (\text{relativistic}) \ . \label{eq:EinsteinR}
\end{equation}

To see how the Einstein relation fixes the correct non-relativistic equilibrium distribution, one can consider the non-relativistic limit and make $B$ constant and equal to Eq.~\eqref{eq:EinsteinNR}. The FP equation for $p=|\bm{p}|$ is
\begin{equation}
\frac{\partial f (p,t)}{\partial t} = \frac{1}{p^2} \frac{\partial}{\partial p} \left[ p^3 \left(A f(p,t) + \frac{AmT}{p} \frac{\partial f(p,t)}{\partial p} \right) \right] \ ,
\label{eq:FP3Dmodulus}
\end{equation}
and the stationary solution is obtained from
\begin{equation}
    f_{\text{eq}}(p)+ \frac{mT}{p} \frac{\partial f_{\text{eq}}(p)}{\partial p} =0 \ , \quad (\text{non-relativistic})
\end{equation}
whose solution is the Maxwell--Boltzmann distribution function,
\begin{align}
f_{\text{eq}}(p) & \propto \exp\left( -\frac{p^2}{2mT} \right) \ . \quad (\text{non-relativistic})
\label{eq:MB_nr}  
\end{align}

In the relativistic case, for constant $A$, and energy-dependent $B$ given by Eq.~\eqref{eq:EinsteinR}, one obtains the stationary solution from,
\begin{equation}
    f_{\text{eq}}(p)+ \frac{T}{p} \frac{\partial [ (E+T) f_{\text{eq}}(p)]}{\partial p} =0 \ , \quad (\text{relativistic})
\end{equation}
whose solution is the  J\"uttner distribution
\begin{align}
f_{\text{eq}}(p) & \propto \exp\left( -\frac{E}{T} \right) \ . \quad (\text{relativistic})
\label{eq:MB_r}  
\end{align}

The normalization of the distribution function $f(\bm{p},t)$ is chosen so that 
\be N= \int_{ \mathbb{R}^3} \dd^3p \ f(\bm{p},t) \ee
is equal to the  total number of HQs in the system.
In the isotropic case, it is also useful to define the density 
\be n(p,t)=4 \pi p^2 f(p,t) \ , \label{eq:np}  \ee
so that
\be N = \int_0^\infty \dd p \ n(p,t) =\int_0^\infty  \dd p \ \frac{\dd N}{\dd p} (p,t)\ . \ee

In the non-relativistic case, the solution of the FP equation for $n(p,t)$ subject to the initial condition,
\be n(p,t=0) = N \delta(p-p_0) \ , \label{eq:icNR} \ee
is~\cite{fox1978gaussian}
\be 
n(p,t) = \frac{Np}{\sqrt{2 \pi} p_0 \mu(t) \sigma (t)} \left[ e^{-\frac{(p-p_0 \mu(t))^2}{2 \sigma^2(t)}} - e^{-\frac{(p+p_0 \mu(t))^2}{2 \sigma^2(t)}}\right] \ , \label{eq:FPsolution}
\ee
with $\mu(t)=\exp(-At)$ and $\sigma^2(t)=mT[1-\mu^2(t)]$. Notice that for $t\rightarrow\infty$, it is consistent with Eq.~\eqref{eq:MB_nr}. In the relativistic case, no analytical solution of the FP equation is known.

We summarize in Table~\ref{tab:param} the parameters we use in this work for the non-relativistic and relativistic calculations, respectively. As mentioned, we use, for simplicity, a constant drag coefficient $A$. The values for the friction coefficient are calculated from the fluctuation-dissipation theorem of Eq.~\eqref{eq:fricforce}.

\begin{table}[ht]
    \centering
     \begin{tabular}{|c||c|c|c|}
     \hline
          & Drag coeff. & Diffusion coeff. & Friction coeff.\\
          \hline
         NR & $A$ & $B=AmT$ & $\Gamma = A$ \\
         R & $A$ & $B(E)=A(E+T)T$ & $\Gamma(E)=A + AT/(2E)$ \\
      \hline
    \end{tabular}
    \caption{Parameters used in this work in the Langevin and Fokker--Planck equations in the non-relativistic (NR) and relativistic (R) cases. The coefficient $A$ is always kept as a constant. We use $A=0.4$ fm$^{-1}$ and $A=0.2$ fm$^{-1}$ in the NR and R cases, respectively.}
    \label{tab:param}
\end{table}

\subsection{Equilibrium current--current correlation function: non-relativistic limit}

In this section, we focus on the non-relativistic limit  of the Langevin equation with white noise, Eq.~\eqref{eq:langevin}. In this limit, it simplifies to
\begin{empheq}[left=\empheqlbrace]{align}
\frac{\dd x^i}{\dd t} &= \frac{p^i}{m} \\
\frac{\dd p^i }{\dd t} & = -A p^i  + \eta^i (t) \ , \label{eq:langevin2}
\end{empheq}
since $E \simeq m$ and $\Gamma=A$ according to the fluctuation-dissipation relation. We consider the equilibrium situation in a bath of temperature $T$, with $T \ll m$.

The momentum correlation function of the HQ, $C_p(t)$, is defined as
\begin{equation}
    C_p (t) \equiv \langle \bm{p} (t) \cdot \bm{p} (0) \rangle \ . \label{eq:Cp}
\end{equation}

From Eq.~\eqref{eq:langevin2} this quantity satisfies the deterministic equation
\begin{equation}
 \frac{ \dd C_p(t)}{\dd t }=- A C_p(t) \ , 
\end{equation}
whose solution is an exponential decay,
\begin{equation}
    C_p(t)=C_p(0) \ e^{-At} \ , \label{eq:simplerelax}
\end{equation} 
with $C_p(0)=\langle \bm{p} (0) \cdot \bm{p} (0)\rangle = 3Tm$, and a relaxation time which is the inverse of the drag coefficient,
\begin{equation}
    \tau_R \equiv A^{-1} \ .
\end{equation}

From a more macroscopic point of view, a swarm of heavy particles defines a local current $\bm{J} (\bm{x},t)$. This current is globally conserved but not locally. Imposing spatial homogeneity of the system, we define the volume-averaged current~\cite{Hammelmann:2018ath},
\begin{equation}
j^i(t) \equiv \frac{1}{V}\int \dd^3x \, J^i(\bm{x},t) \ , 
\label{eq:mesoscopic_current}
\end{equation}
where the volume $V$ could be the entire volume of the system, or a finite subvolume. This current presents fluctuations due to the random motion of HQs in it, whose intensity decreases with $V$.

Microscopically, this fluctuating current is related to the momentum of the heavy particles. In the non-relativistic domain,
\begin{equation}
    \bm{j}(t)= \frac{q}{Vm} \sum_a {\bm p}_a(t) \ , \label{eq:NRcurrent}
\end{equation}
where the sum runs over all particles ($a=1,2...,N$) contained within $V$. $q$ is the charge of the particles, which we just set to $q=1$ (HQ charge).

The current--current correlation function (see App.~\ref{app:hydrofluc} for an extended discussion within the theory of hydrodynamic fluctuations) is defined as,
\begin{equation}
C(t) \equiv \langle \bm{j} (t) \cdot \bm{j} (0) \rangle \ ,
\end{equation}
and inherits the same exponential decay of $C_p(t)$,
\begin{equation}
    C(t) = \frac{q n_c}{Vm^2} C_p(t)= C(0) e^{-A t} \ , \label{eq:Ctmemoryless}
\end{equation}
where
\begin{equation}
    C(0)=\frac{3T q n_c}{Vm} \ , \label{eq:C0NR}
\end{equation}
and $n_c$ is the charge density
\begin{equation}
    n_c= \frac{qN}{V} = \frac{q}{V} \int \frac{\dd^3p}{(2\pi)^3} f_{\text{eq}}(\bm{p}) \ .
\end{equation}
In the non-relativistic case we can explicitly compute it as:
\begin{equation}
    n_c=  q z \left( \frac{Tm}{2\pi} \right)^{3/2} e^{-m/T} \ , \label{eq:densityNR}
\end{equation}
where $z$ is the fugacity,
\begin{equation}
    z =\exp \left( \frac{q\tilde{\mu}}{T} \right) \ ,
\end{equation}
and $\tilde{\mu}$ the HQ chemical potential.

Using the explicit form of the current--current correlation function~\eqref{eq:Ctmemoryless}, we can straightforwardly obtain
\begin{equation}
\int_0^\infty \dd t \ C(t)= \frac{C(0)}{A}
= \frac{3T^2 \tilde{\chi}}{VmA} \ , \label{eq:intCt}
\end{equation}
where, for convenience, we have introduced the susceptibility (see App.~\ref{app:hydrofluc}),
\be \tilde{\chi} = \left( \frac{\partial n_c}{\partial \tilde{\mu}} \right)_T = \frac{q n_c}{T} \ . \ee

On the other hand, we introduce the spatial diffusion coefficient $D_s$, as the relation between the mean squared displacement of the Brownian particle and time,
\be \langle [{\bm r} - {\bm r(t=0)}]^2 \rangle \simeq 6 D_s t \ . \ee
It can be directly related to the momentum diffusion coefficient in the overdamped limit of the non-relativistic Langevin equation~\cite{landau1981physical} , 
\be D_s = \frac{T^2}{B} \ . \label{eq:DsB} \ee
Using the Einstein relation Eq.~(\ref{eq:EinsteinNR}) one can also relate it to the drag coefficient,
\begin{equation}
    D_s = \frac{T}{mA} \ . \label{eq:DsA}
\end{equation}

Now, introducing Eq.~(\ref{eq:DsA}) into the result (\ref{eq:intCt}) we can write
\begin{equation}
D_s =
\frac{V}{3 T \tilde{\chi}}
\int_0^{\infty} \dd t\,
\langle \bm{j}(t)\!\cdot\!\bm{j}(0) \rangle \ .
\label{eq:GK_Ds}
\end{equation}
This expression coincides with the Green--Kubo formula~\cite{landau1981physical,toda1992statistical,zubarev1996statistical}, relating a transport coefficient to a correlation function of a conserved hydrodynamic field. The Green--Kubo formula, derived within linear response theory, is generally valid for any form of the correlation function. Here, it is a consequence of the Langevin equation, whose delta-correlated noise is explicitly known (it is also derived in App.~\ref{app:hydrofluc} using the theory of hydrodynamic fluctuations for the exponentially-correlated noise case). 

The expression given in Eq.~\eqref{eq:GK_Ds} is used to test the equilibrium current--current correlation function and to verify the self-consistency of the approach by calculating $D_s$ from the microscopic dynamics. This has been applied in box calculations to address other transport coefficients, e.g. in Refs.~\cite{Muronga:2003tb,Demir:2008tr,Wesp:2011yy,Plumari:2012ep,Rose:2017bjz,Hammelmann:2018ath,Rose:2020lfc,Hammelmann:2023fqw}.

\subsection{Correlation function in the relativistic case \label{sec:GK_relativistic}}

When we depart from the non-relativistic dynamics, the previous derivation cannot be easily adapted since the term $\dd \bm{p}/\dd t$ is not proportional to the momentum (the friction coefficient now depends on energy); and the HQ current is also not proportional to the HQ momenta as in Eq. \eqref{eq:NRcurrent}. Then it is not possible to easily connect and find a closed expression for either $C_p(t)$ or $C(t)$.

In the relativistic case, we still use the Green--Kubo formula given in Eq.~\eqref{eq:GK_Ds} to compute $D_s$, but assuming an exponential decay of $C(t)$, which has also been shown to be valid in the previous literature~\cite{Muronga:2003tb,Demir:2008tr,Wesp:2011yy,Plumari:2012ep,Rose:2017bjz}. The exponential form has been explicitly checked for all temperatures considered here. 

Therefore, we take
\begin{equation}
C(t)=C(0) \ e^{ -t/\tau} \ , \label{eq:Ctexp}
\end{equation}
where $\tau$ is a fitting parameter of the correlation function. We do not assume any \textit{a priori} relation to any physical quantity, but we have in mind that it should be close to $1/\Gamma(\langle E \rangle)$, and that in the non-relativistic limit it should coincide with $\tau_R$, as we have demonstrated before.

The equal-time correlation function $C(0)$ is an equilibrium property of the medium, different from Eq.~\eqref{eq:C0NR}. It can nevertheless be calculated in the relativistic case as,
\begin{equation}
    C(0)=\frac{q^2}{V} \int \frac{\dd^3p}{(2\pi)^3} \frac{p^2}{E_p^2} f_{\text{eq}}(\bm{p}) \ . \label{eq:C0memoryless}
\end{equation}
Introducing the J\"uttner distribution, we obtain
\begin{equation}
 C(0)= \frac{zq^2T^3}{2 \pi^2 V} {\cal I} (y) \ , \label{eq:C0rel}
\end{equation}
with $y=m/T$. The dimensionless ${\cal I}(y)$ function reads,
\begin{equation}
{\cal I} (y)= y^3 \int_1^\infty \dd w w^{-1} (w^2-1)^{3/2} e^{-yw} \ ,
\end{equation}
where $w=E/m$.

Notice that in the relativistic domain, $C(0)$ is not anymore proportional to the HQ density, $n_c$, but to a different quantity. However, in the non-relativistic limit $y \gg 1$, one has ${\cal I} (y) \rightarrow 3 \sqrt{\pi/2} \ y^{1/2} \exp(-y)$, and $C(0)$ goes to the non-relativistic value in Eq.~\eqref{eq:C0NR}. 

The HQ density in the relativistic domain can also be computed using the Jüttner distribution and we obtain
\begin{equation}
    n_c= \frac{qzT^3}{2\pi^2} {\cal J} (y) \ ,
\end{equation}
where the ${\cal J}(y)$ function reads
\begin{equation}
{\cal J} (y)=y^3 \int_1^\infty \dd w \ w \sqrt{w^2-1} \ e^{-yw} \ .
\end{equation}
In a more standard fashion, one can relate this function with the modified Bessel function of the second kind, ${\cal J}(y)=y^2 K_2(y)$. As an additional check, in the non-relativistic limit ($y \gg 1$) one immediately recovers the expression of Eq.~\eqref{eq:densityNR}.

Now, inserting the $C(t)$ in Eq.~\eqref{eq:Ctexp} into the Green--Kubo expression, we get
\be D_s = \frac{C(0) V \tau}{3 T \tilde{\chi}} \ , \label{eq:Dsrel} \ee
which allows us to obtain the diffusion coefficient in the relativistic domain, once we fit $C(t)$ to the exponential form given in Eq.~\eqref{eq:Ctexp} and extract the values of $C(0)$ and $\tau$.

We insist that in the relativistic case, we cannot find closed relations between $D_s$ and $A$ (or $B$) due to the lack of exact expressions for the correlation function. To our knowledge, such expressions are not known~\cite{dunkel2005theory}. In fact, we have only found expressions for the one-dimensional case, for different stochastic prescriptions, and for a constant $A$ coefficient (like our setup), in Ref.~\cite{Dunkel:2008ngc}.

Nevertheless, we can make an educated guess using Eq.~\eqref{eq:DsB} and exploiting the relativistic version of the Einstein relation in Eq.~\eqref{eq:EinsteinR}. We get
\begin{equation}
D_s = \frac{T}{A ( \langle E \rangle + T)} \ ,
\label{eq:einstein_relation}
\end{equation}
where the average thermal energy reads~\cite{greiner2012thermodynamics},
\be \frac{\langle E \rangle}{T} = 3 + y \frac{K_1(y)}{K_2(y)} \ , \label{eq:avE} \ee
which describes both the non-relativistic and ultrarelativistic limits, respectively, $\langle E \rangle = m + 3T/2 + \cdots$ and $\langle E \rangle = 3T + \cdots$.

In the non-relativistic limit $m \gg T$, Eq.~\eqref{eq:einstein_relation} takes the correct value $D_s=T/(mA)$, but we insist that, in the relativistic domain, Eq.~\eqref{eq:einstein_relation} is just used as a guess for the spatial diffusion coefficient in terms of the drag coefficient.

\section{Colored noise and the generalized Langevin equation}
\label{sec:colored_noise}

In realistic many-body systems, the response of macroscopic currents to external perturbations is not instantaneous. 
Microscopic interactions introduce a finite time scale over which the system retains memory of its past evolution.
This leads to non-Markovian dynamics characterized by colored noise and generalized Langevin equations~\cite{fox1977generalized,fox1978gaussian,haunggi1994colored}.
Such effects might be relevant for HQ transport in the QGP~\cite{Ruggieri:2019zos,Ruggieri:2022kxv}, where microscopic scattering processes occur on time scales comparable to the relaxation time of the medium.

In this section, we formulate the generalized Langevin and Fokker--Planck equations with memory, discuss their equilibration dynamics, and derive the corresponding current--current correlation functions, emphasizing their relation to transport coefficients. As in the white noise case, we are only able to obtain analytical results in the non-relativistic domain of the generalized Langevin and Fokker--Planck equations.

\subsection{Generalized Langevin equation and equilibration}
\label{subsec:equilibration_memory}

We consider the generalized Langevin equation for the HQ~\cite{Ruggieri:2022kxv}, in which a memory kernel is added to the deterministic term,
\begin{equation}
\frac{ \dd \bm{p} (t)}{\dd t}
=
- \int_0^t \dd s  \, \bm{\gamma}(t-s)\, \bm{p} (s)
+ \bm{\eta} (t) \ ,
\label{eq:GLE}
\end{equation}
where $\bm{\gamma}(t-s)=\gamma^{ij}(t-s)$ is the nonlocal memory kernel that encodes the delayed response of the medium to the HQ motion. In Eq.~\eqref{eq:GLE}, $\boldsymbol{\eta}(t)$ is the stochastic noise, whose correlator is
\begin{equation}
\langle \eta^i(t)\eta^j(s) \rangle
=
2 B(E) f(t-s)\, \delta^{ij} \ ,
\label{eq:noise_colored_general}
\end{equation}
where $f(t - s)$ is a function that characterizes the temporal correlations of the random force. According to the fluctuation--dissipation theorem, the memory kernel $\bm{\gamma}(t-s)$, which characterizes the system's response to external perturbations, is related to the autocorrelation function of $\bm{\eta}(t)$. In the non-relativistic limit we can write for a single space component $\eta(t)$,
\begin{equation}
\bm{\gamma}(t - s) = \frac{1}{mT}\,\langle \eta^i(t)\,\eta^j(s) \rangle = 2 \Gamma f(t - s)\delta^{ij} \ ,
\label{eq:FDT_general}
\end{equation}
where $\Gamma$ is the friction coefficient, which in the non-relativistic limit is equal to $A=B/(mT)$ (cf. Table~\ref{tab:param}).

From Eq.~\eqref{eq:GLE}, the standard (memoryless) Langevin equation can be recovered by considering the limit in which the temporal correlations of the stochastic force in Eq.~\eqref{eq:noise_colored_general} are negligible. In that case,
\begin{equation}
f(t-s) \rightarrow \delta(t-s) \ ,
\end{equation}
and
\begin{equation}
\gamma^{ij}(t - s) =  2 \Gamma \delta(t - s) \delta^{ij}\  .
\label{eq:FDT_general_2}
\end{equation}

In this limit, Eq.~\eqref{eq:GLE} simplifies to Eq.~\eqref{eq:langevin} provided that we use
\begin{equation}
\int_0^t \dd s \, \delta(t-s) = \frac12 \ .
\end{equation}

The numerical implementation of colored noise in the Langevin equation for HQs is much more complicated than the standard white noise case. To generate such a correlated noise, we introduce an ancillary stochastic process~\cite{Ruggieri:2022kxv}, $h(t)$, which evolves simultaneously with, but independently of, the HQs. It is constructed so that its time-correlator remains non-vanishing.

We begin with a stochastic process (in one spatial dimension, for simplicity) $h(t)$, with $h(t=0)=0$, which obeys the Langevin equation,
\begin{equation}
\frac{\dd h(t)}{\dd t} = -\alpha h(t) + \alpha \xi (t)\ ,
\label{eq:h_ancillary}
\end{equation}
where \(\xi (t)\) is a Gaussian white noise with vanishing mean,
\begin{equation}
\langle \xi(t) \rangle = 0 \ ,
\end{equation}
and two-point correlation,
\begin{equation}
\langle \xi(t)\xi(t') \rangle = \frac{1}{\alpha}\,\delta(t - t') \ .
\end{equation}

Here, $h(t)$ is the ancillary process, introduced as an auxiliary stochastic variable to generate the colored noise for the Langevin equation governing the HQs.

From the definition of $h(t)$, it is straightforward to see that \(\langle h(t) \rangle = 0\). However, its time correlator at different times is not a \(\delta\)-function. Rather, it can be shown that~\cite{Ruggieri:2022kxv},
\begin{equation}
\langle h(t) h(s) \rangle = \frac{1}{2}\left(e^{-\alpha |t - s|} - e^{-\alpha (t + s)}\right).
\label{eq:h_correlator}
\end{equation}

Hence, $h (t)$ is a memory-bearing process. The term $e^{-\alpha (t+s)}$ explicitly depends on the sum of the times rather than their difference. Therefore, it breaks time-translational invariance and represents a non-stationary contribution. However, from Eq.~\eqref{eq:h_correlator}, one asymptotically has the following,
\begin{equation}
\langle h(t)h(s) \rangle \simeq \frac{1}{2} e^{-\alpha |t-s|} \ ,
\label{eq:h_correlator_2}
\end{equation}
i.e., correlations decay over the time scale \( \tau_m \equiv 1/\alpha \). 
Furthermore, in the limit \(\alpha \to +\infty\), Eq.~\eqref{eq:h_correlator}  yields
\begin{equation}
\alpha\langle h(t)h(s) \rangle \simeq \delta(t-s) \ ,
\end{equation}
showing that the process \(h(t)\) reduces to standard white noise in the limit \(\tau_m \to 0\), as expected.

It is important to note that the correlations of \(h (t)\) develop over a time scale \(t \sim \tau_m\), time required to suppress the non-stationary term \(e^{-\alpha (t+s)}\) of Eq.~\eqref{eq:h_correlator}. After this transient regime, the time correlations are well approximated by Eq.~\eqref{eq:h_correlator_2} and become translationally invariant. Hence, the noise becomes approximately stationary. 

In addition, the steady-state variance of \(h (t)\) is given by
\begin{equation}
\langle h^2(t) \rangle = \frac{1}{2}\left(1 - e^{-2\alpha t}\right),
\end{equation}
which approaches
\begin{equation}
\langle h^2 (t)\rangle = \frac{1}{2}
\end{equation}
at late times. This limit is reached for \(t \gtrsim \tau_m = 1/\alpha\), where the term \(e^{-2\alpha t}\) is exponentially suppressed. Hence, the process in Eq.~\eqref{eq:h_ancillary} describes a noise that requires a time \(\sim \tau_m\) or longer to develop memory (i.e., correlations).

In the generalized Langevin equation~\eqref{eq:GLE}, we can express the correlated noise \(\eta(t)\) in terms of \( h(t)\), ensuring that \(\eta\) reproduces the correct delta-correlated limit for white noise. This yields
\begin{equation}
\eta(t) = \sqrt{\frac{2B}{\tau_m}} \, h(t) \ . 
\end{equation}

With this choice, our nonlocal memory kernel is an exponentially decaying function,
\begin{equation}
\gamma(t-s) = \frac{A}{\tau_m} e^{-|t-s|/\tau_m} ,
\label{eq:exp_kernel}
\end{equation}
where $\tau_m$ defines the memory time of the noise.
In the previous equation we have used that $A=\Gamma$ in the non-relativistic limit. However, in the relativistic case, as known from previous sections, they differ in the Stratonovich-Fisk prescription and, in fact, $\Gamma(E)$ is energy dependent even when $A$ is constant.

Using Eq.~\eqref{eq:exp_kernel}, the generalized Langevin equation becomes
\begin{equation}
\frac{\dd {\bm{p}}(t)}{\dd t}
=
- \frac{A}{\tau_m}
\int_0^t \dd s \ e^{-\frac{t-s}{\tau_m}} \ \bm{p}(s)
+ \boldsymbol{\eta}(t) \ ,
\label{eq:GLE_exp}
\end{equation}
with $\langle \eta^i\rangle=0$ and
\be
\langle \eta^i(t) \eta^j(s) \rangle
=
\frac{B}{\tau_m} e^{-\frac{|t-s|}{\tau_m}}\, \delta^{ij} \  . \ee

Despite the non-Markovian structure, equilibration is guaranteed~\cite{fox1978gaussian, Greiner:1998vd}. At late times, the momentum distribution relaxes toward the thermal equilibrium form, which satisfies 
\begin{equation}
\langle \bm{p}^2 \rangle = 3 T m \ ,
\label{eq:thermal_equilibrium}
\end{equation}
provided that the fluctuation--dissipation relation is satisfied.

To conclude this section, we try to understand the impact of the memory time as a delayed response due to the presence of the memory kernel. Let us focus on the deterministic part of Eq.~\eqref{eq:GLE_exp},
\begin{equation}
\frac{ \dd \langle \bm{p} (t) \rangle}{\dd t}
=
- \frac{A}{\tau_m} \int_0^t \dd s \  e^{-\frac{t - s}{\tau_m}}  \, \langle \bm{p}(s) \rangle \ .
\label{eq:GLE_DExp}
\end{equation}
In this equation of motion for $\langle \bm{p}(t) \rangle$, we differentiate with respect to time once and apply the Leibniz rule. We get the differential equation,
\begin{equation}
\frac{\dd^2 \langle \bm{p}(t) \rangle}{\dd t^2} + \frac{1}{\tau_m}\frac{\dd \langle \bm{p} (t) \rangle}{\dd t} + \frac{A}{\tau_m} \langle \bm{p}(t) \rangle = 0 \ ,
\label{eq:DHO}
\end{equation}
which takes the same form as the equation of a damped harmonic oscillator, with damping coefficient $1/\tau_m$ and frequency $\sqrt{A/\tau_m}$. The condition for having oscillations of $\langle \bm{p}(t) \rangle$ is $4 A \tau_m > 1$. Otherwise, the dynamics are overdamped and no oscillations occur.

It becomes clear that in the presence of memory, the rate of momentum change cannot adjust instantaneously, as memory introduces a time delay in the system’s response. This delayed response effectively generates second-order dynamics, which behaves like inertia.

The introduction of noise complicates the analysis without altering these fundamental features. With noise, one can obtain,
\begin{equation}
\frac{\dd^2 p(t)}{\dd t^2} + \frac{1}{\tau_m}\frac{\dd p(t)}{\dd t} + \frac{A}{\tau_m} p(t) = \frac{\dd \eta(t)}{\dd t} + \frac{1}{\tau_m}\eta(t) \ .
\label{eq:DHO_N}
\end{equation}
Even in the presence of noise, we obtain the same damped oscillator structure. The noise does not eliminate the oscillations; rather, it acts as a driving term that perturbs and modulates them.

Concerning the relativistic case, we insist that apart from the modifications in kinematics, the diffusion coefficient $B(E)$ and the friction coefficient $\Gamma(E)$ are energy dependent for constant $A$. The numerical implementation of this case in the Stratonovich-Fisk prescription is similar to the case of white noise, since the ancillary process $h(t)$ is not affected.

\subsection{Correlation function in the presence of memory}
\label{subsec:current_correlation_memory}

To quantify the impact of memory effects on transport coefficients, we first study the momentum correlation function $C_p(t)$ defined in Eq.~\eqref{eq:Cp}. In this section, we work in the non-relativistic domain.

The equation of motion of $C_p(t)$ can be constructed using the generalized Langevin equation~\eqref{eq:GLE_exp}. After multiplying by $\bm{p}(0)$ and taking averages, one finds
\begin{equation}
\frac{\dd C_p(t)}{\dd t}
= - \frac{A}{\tau_m}
\int_0^t \dd s \ e^{-\frac{t-s}{\tau_m}} \ C_p(s) \ .
\label{eq:Cp_eom}
\end{equation}

This equation is formally similar to Eq.~\eqref{eq:GLE_DExp}, which led to the damped solution for $\langle \bm{p}(t) \rangle$. To solve Eq.~\eqref{eq:Cp_eom}, we employ the Laplace transform of the time correlation function,
\begin{equation}
\tilde{C}_p(s) \equiv \int_0^{\infty} \dd t \ e^{-st} C_p(t) \ .
\end{equation}
Using the convolution property, it reads
\begin{equation}
\tilde{C}_p(s)
= C_p(0)
\frac{1+\tau_ms}
{\tau_m s^2 + s + A} \ . 
\label{eq:Cp_laplace}
\end{equation}

The denominator of Eq.~\eqref{eq:Cp_laplace} has two roots,
\begin{equation}
s_{\pm}
\equiv
\frac{-1 \pm \sqrt{\Delta}}{2\tau_m} \ ,
\label{eq:roots}
\end{equation}
where
\begin{equation}
     \Delta \equiv 1 - 4 A \tau_m = 1- 4 \ \frac{\tau_m}{\tau_R}\ .
\end{equation}

Depending on the sign of $\Delta$, the system exhibits either overdamped or underdamped behavior. To see this, we take the inverse Laplace transform of Eq.~\eqref{eq:Cp_laplace} and get
\begin{equation}
C_p(t)
= C_p(0) \ \chi(t) \ , \label{eq:Cpmemory}
\end{equation}
with the function $\chi(t)$ defined as the sum of two exponentials,
\begin{equation}
    \chi(t) \equiv \frac{ s_+ e^{s_- t}
- s_- e^{s_+ t}}
{s_+ - s_-} \ . \label{eq:chit}
\end{equation}
Notice that when $\tau_m \rightarrow 0$, then $\chi(t) \rightarrow \exp(-At)$ and one recovers the white noise results of Eq.~\eqref{eq:simplerelax}.

When $\Delta >0$ (overdamped case) the two exponentials are real and one can further write
\be C_p(t)= C_p(0) e^{-\frac{t}{2 \tau_m}} \left[ \cosh \left( \frac{\sqrt{\Delta} t}{2 \tau_m} \right) + \frac{1}{\sqrt{\Delta}}  \sinh \left( \frac{\sqrt{\Delta} t}{2 \tau_m} \right) \  \right] \ . 
\label{eq:CpmemoryOD}\ee

If $\Delta <0$ (underdamped limit) the exponentials are complex and $s_\pm$ are complex conjugate roots. The correlation function can be written,
\be C_p(t)=C_p(0) e^{-\frac{t}{2\tau_m}} \left[ \cos \left(  \frac{\sqrt{|\Delta|} t}{2 \tau_m} \right) + \frac{1}{\sqrt{|\Delta|}} \sin \left( \frac{\sqrt{|\Delta|} t}{2 \tau_m} \right) \right] \ .
\label{eq:CpmemoryUD}
\ee

A key nontrivial result is that the time integral of the correlation function is independent of the memory time. We find
\begin{equation}
\int_0^{\infty} \dd t \ C_p(t)
=
\frac{C_p(0)}{A} \ ,
\label{eq:Cp_integral}
\end{equation}
which can be checked explicitly in any of Eqs.~\eqref{eq:Cpmemory}, \eqref{eq:CpmemoryOD} or \eqref{eq:CpmemoryUD}.

This important result implies that the corresponding current--current correlator,
\begin{equation}
C(t)=\langle \bm{j}(t)\cdot \bm{j}(0) \rangle
=
\frac{q n_c}{V m^2} C_p(t) \ , 
\label{eq:current_corr_memory}
\end{equation}
leads again to the Green--Kubo relation
\begin{equation}
\frac{V}{3T\tilde{\chi}}
\int_0^{\infty} \dd t \,
C(t)
=
\frac{C_p(0)}{3m^2 A} = \frac{T}{m A} = D_s \ .
\label{eq:GK_memory}
\end{equation}

Equation~\eqref{eq:GK_memory} demonstrates that, while memory effects qualitatively modify the temporal structure of current correlations, the value of the spatial diffusion coefficient remains unchanged in the non-relativistic limit, when the correlation function is integrated up to $t=+\infty$, which is the genuine hydrodynamic limit.

Similarly to the white noise case, the analytical results obtained cannot be extended to the relativistic case. However, in analogy to that case, we use the functional forms of the non-relativistic correlation functions (\ref{eq:CpmemoryOD}) and (\ref{eq:CpmemoryUD}) to fit those in the relativistic domain, checking case by case that it is sensible to do that.  

The combination of the relativistic case (multiplicative noise) with correlated noise raises an additional issue. The limit $\tau_m \rightarrow 0$ and the continuous matching to the Markovian, uncorrelated noise limit is delicate. According to the Wong-Zakai theorem~\cite{wong1969riemann,sussmann1978gap,van1983stochastic,oksendal2003stochastic}, the limit only makes sense when the Stratonovich-Fisk discretization prescription is used. Possibly related to this fact, we found small deviations in the expected thermal equilibrium distribution when the Itô prescription was applied to the case with a memory kernel. For these reasons, we adopt the Stratonovich-Fisk prescription in this work, where we checked that such deviations no longer appear.

\subsection{Generalized Fokker--Planck equation~\label{subsec:gener_FP}}

We consider now the mesoscopic description of the system, and work with the time evolution of the HQ distribution function in the presence of memory. For this, we need a non-Markovian generalization of the FP equation, equivalent to the generalized Langevin equation.

The generalized FP equation in the non-relativistic limit reads~\cite{fox1977generalized,fox1978gaussian},
\begin{equation}
\frac{\partial f (\bm{p},t)}{\partial t} = \frac{\partial}{\partial p^i} \left[ -\frac{\dot{\chi}(t)}{\chi(t)}  p^i f(\bm{p},t)  -\frac{\dot{\chi}(t)}{\chi(t)}   T m \frac{\partial  f(\bm{p},t)}{\partial p^i} \right] \ ,
\label{eq:GFP3D}
\end{equation}
where $\chi(t)$ is defined in Eq.~\eqref{eq:chit}. This equation is Gaussian but non-Markovian~\cite{fox1978gaussian} in line with the correlated noise appearing in its stochastic Langevin counterpart.

As in the Markovian FP case, the solution of this equation in time has an analytical solution in the non-relativistic case. However, the same is not the case in the relativistic domain. As before, we consider the solution for the distribution function in the isotropic 3D case, as a function of the modulus of $\bm{p}$. We consider the HQ density $n(p,t)$ as defined in Eq.~\eqref{eq:np}. 

Subject to the initial condition $n(p,0) = N \delta(p-p_0)$, the Gaussian solution we find is rather similar to the FP case. It reads~\cite{fox1978gaussian}
\be 
n(p,t) = \frac{Np}{\sqrt{2 \pi} p_0 \chi(t) \sigma (t)} \left[ e^{-\frac{(p-p_0 \chi(t))^2}{2 \sigma^2(t)}} - e^{-\frac{(p+p_0 \chi(t))^2}{2 \sigma^2(t)}}\right] \ ,
\ee
where the function $\chi(t)$ is given in Eq.~\eqref{eq:chit} and the variance is $\sigma^2(t)=mT[1-\chi^2(t)]$. In the $\tau_m \rightarrow 0$ limit, $\chi(t) \rightarrow \mu(t)=\exp(-At)$ and one recovers the memoryless solution of Eq.~\eqref{eq:FPsolution}.

While the analytical solution in the relativistic case is not known, we know that the asymptotic solution should still be given by the J\"uttner distribution.

\section{Results}

In this section, we provide a series of numerical results under different conditions: nonequilibrium and equilibrium situations, non-relativistic and relativistic dynamics, white (uncorrelated) and colored (correlated) Gaussian noise. In all cases, we use static box calculations at constant HQ density, $n_c$ and $T$. In addition, we use the mid-point (Stratonovich-Fisk) discretization prescription with a constant value of $A$, and a $B(E)$ given by Table~\ref{tab:param}, depending on the use of relativistic or non-relativistic cases. While not reported here for simplicity, we have also simulated the pre-point and post-point discretization schemes.

\subsection{Nonequilibrium: relaxation to equilibrium}

\subsubsection{Non-relativistic evolution with white noise}

First of all, we test the non-relativistic evolution out of equilibrium using the Langevin and Fokker--Planck equations. For the latter, an analytical solution of the time evolution, with a Dirac delta initial condition, has been given in Eq.~\eqref{eq:FPsolution}.

\begin{figure}[t!]
    \begin{center}
    \includegraphics[width=\linewidth]{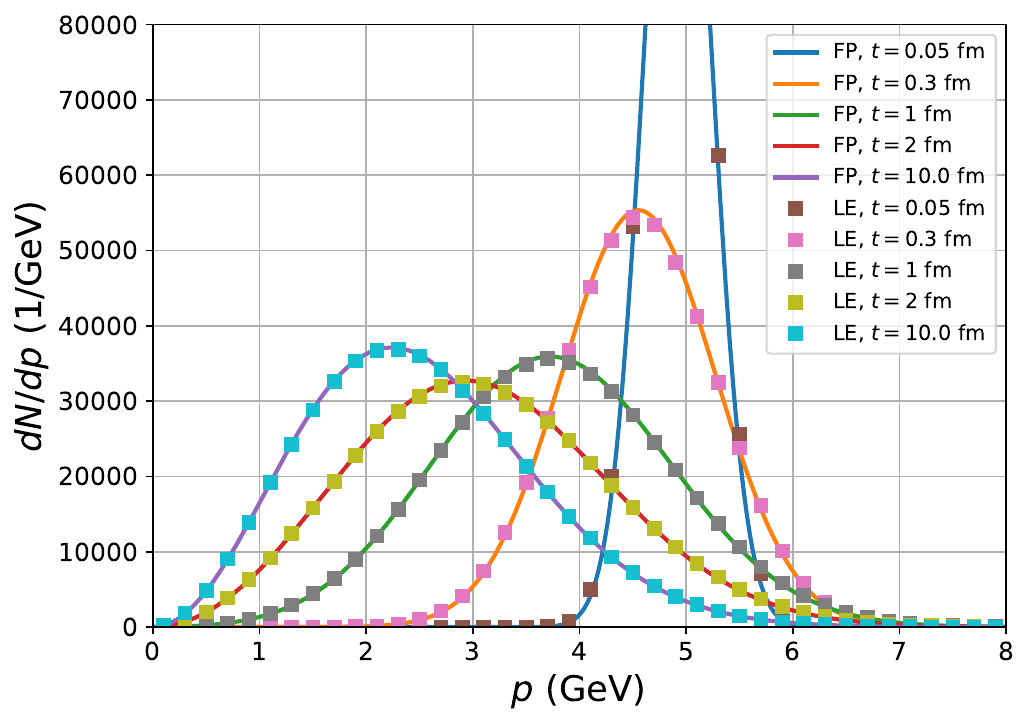}
    \end{center}
    \caption{ Heavy-quark momentum distribution ($\dd N/\dd p$) evolution obtained from numerical Langevin simulations in a non-relativistic system (LE) and the analytical solution of the Fokker--Planck equation (FP). Results are shown for a drag coefficient $A=0.4\,\mathrm{fm}^{-1}$, temperature $T=0.1\,\mathrm{GeV}$, mass $m=25\,\mathrm{GeV}$, an ensemble size of $N=10^5$ and $N_{\textrm{ev}}=50$ events.}
    \label{fig:LEvsFP_NR}
\end{figure}

The simulation runs in a cubic box of size $L=16$ fm, with periodic boundary conditions. The HQ mass is chosen to be $m=25$ GeV and the temperature is $T=0.1$ GeV. We choose a drag coefficient of $A=0.4$ fm$^{-1}$, and a diffusion coefficient given by Eq.~\eqref{eq:EinsteinNR}. The initial out-of-equilibrium HQ distribution is of the form~\eqref{eq:icNR}, with $p_0=5$ GeV. We use $N=10^5$ heavy particles (equivalent to $10^5$ realizations of a single Brownian particle).

We have chosen such a huge, non-physical mass to be able to address the non-relativistic limit dynamically, that is, using the relativistic Langevin equation~\eqref{eq:langevin}. Since $m/T \gg 1$ we expect the results to match well with the analytical solution of the non-relativistic FP equation.

Figure~\ref{fig:LEvsFP_NR} shows the momentum probability distribution $\dd N(p,t)/\dd p=n(p,t)$ of the HQs as a function of time between $t=0.05$ fm and $t=10.0$ fm. The dots represent the Langevin result by averaging over the $N=10^5$ particles, and over $N_{\textrm{ev}}=50$ events. Starting from the initial peaked distribution around $p=5$ GeV the time evolution makes $n(p,t)$ broader (controlled by the diffusion coefficient) and peaked at smaller momentum (controlled by the drag coefficient). The result at $t=10$ fm is already very close to the equilibrium distribution, which we checked analytically. 

The solid lines in Fig.~\ref{fig:LEvsFP_NR} are the analytical solution of the non-relativistic FP Eq.~\eqref{eq:FPsolution}, which coincides very nicely with the numerical solution of the (relativistic) Langevin equation for non-relativistic HQs.

\subsubsection{Relativistic evolution with white noise}

We now consider a more realistic scenario and study the dynamical relaxation of heavy quarks of mass $m=1.27$ GeV, described by the same Langevin Eq.~\eqref{eq:langevin}.
The drag coefficient is reduced to $A = 0.2$ fm$^{-1}$, but the temperature is higher, $T = 0.4$ GeV, and the cubic box has a side $L = 16$ fm. The diffusion coefficient $B(E)$ is now fixed by the Einstein relation, Eq.~\eqref{eq:EinsteinR}. The initial condition is a Dirac delta distribution, peaked at $p_0 = 15$ GeV, as in~\eqref{eq:icNR}. 

\begin{figure}[t!]
    \begin{center}
    \includegraphics[width=\linewidth]{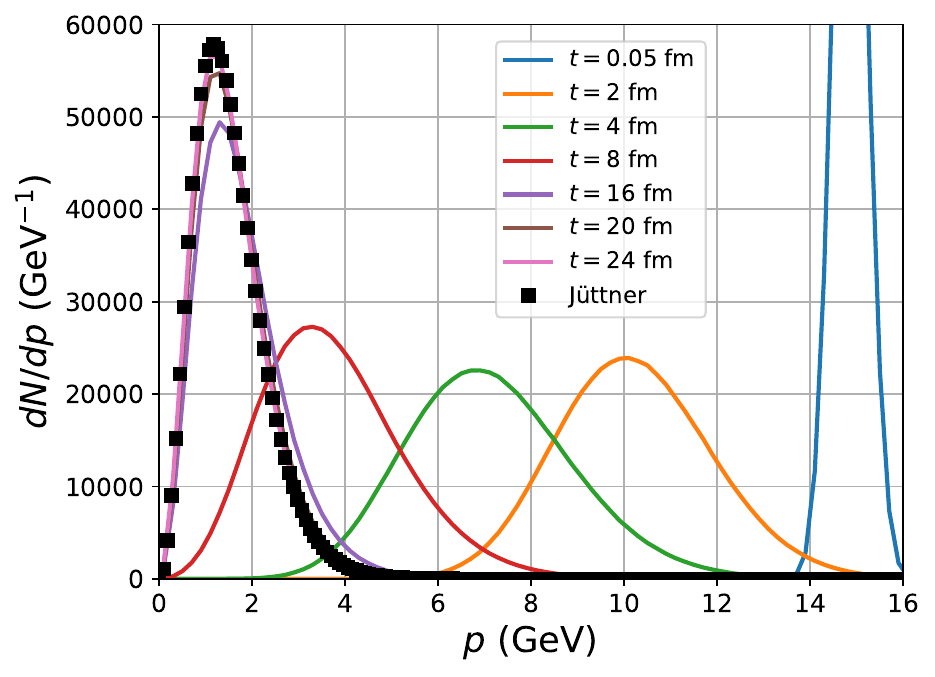}
    \end{center}
    \caption{ Heavy-quark momentum distribution (\(\dd N/\dd p\)) evolution obtained from numerical Langevin simulations in a relativistic system. Results are shown for a drag coefficient \(A=0.2\,\mathrm{fm}^{-1}\), temperature \(T=0.4\,\mathrm{GeV}\), mass \(m=1.27\,\mathrm{GeV}\), an ensemble size of \(N=10^5\), and \(N_{\textrm{ev}}=50\) events.}
    \label{fig:langevin_static_qgp}
\end{figure}

Figure~\ref{fig:langevin_static_qgp} shows the time evolution of the momentum probability distribution $\dd N(p,t)/\dd p=n(p,t)$ obtained from the numerical solution of the Langevin dynamics, again sampled over $N=10^5$ stochastic trajectories and $N_{\textrm{ev}}=50$ events. The figure displays $n(p,t)$ at different times from $t=0.05$ fm to $t = 24$ fm. At $t=0.05$ fm we again observe the initial sharp peak at $p = 15$ GeV. The delta-function initial condition, while idealized, mimics the scenario of a high-momentum parton injected into a quark-gluon plasma, losing energy and thermalizing via elastic collisions. For early times, $t \lesssim 10$ fm, the distribution broadens while its center shifts toward lower momenta. At later times, the distribution continues its Gaussian form, centered around a mean momentum close to the equilibrium one. For $t \gtrsim 20\,\text{fm}$, the distribution is clearly stationary and centered near the expected J\"uttner form, shown in square symbols.

\subsubsection{Non-relativistic evolution with colored noise: exponential memory kernel}

\begin{figure}[t!]
    \begin{center}
    \includegraphics[width=\linewidth]{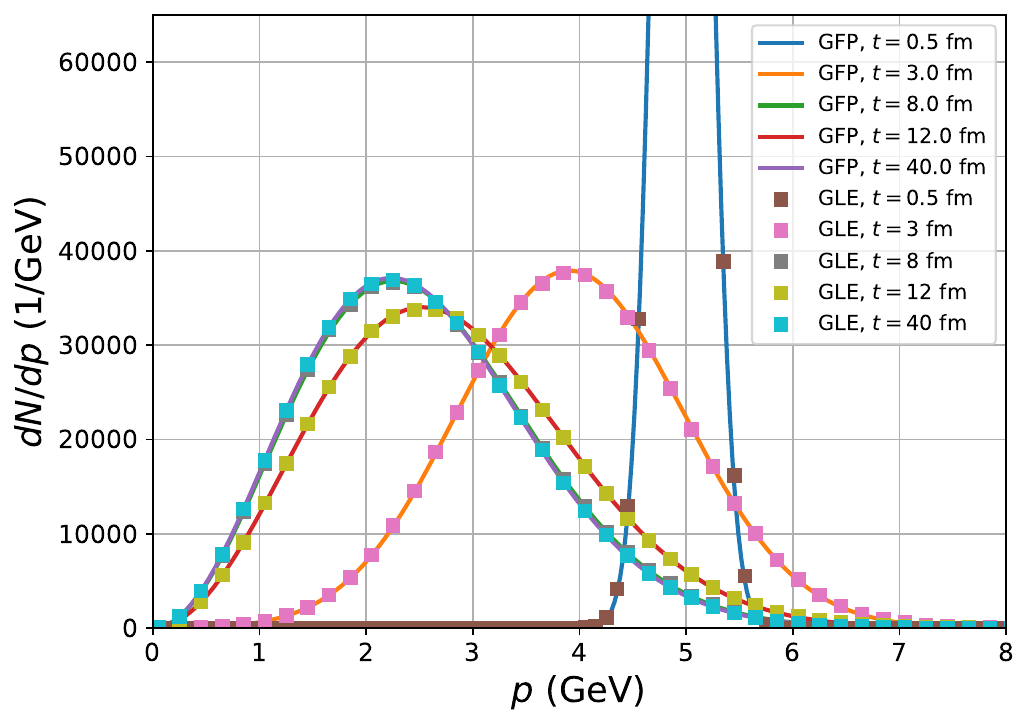}
    \end{center}
    \caption{Heavy-quark momentum distribution (\(\dd N/\dd p\)) evolution obtained from numerical generalized Langevin simulations in a non-relativistic system (GLE) and the analytical solution of the generalized Fokker--Planck equation (GFP). Results are shown for a drag coefficient \(A=0.4\,\mathrm{fm}^{-1}\), temperature \(T=0.1\,\mathrm{GeV}\), mass \(m=25\,\mathrm{GeV}\), memory time \(\tau_m=5\,\mathrm{fm}\), an ensemble size of $N=10^5$ and $N_{\textrm{ev}}=50$ events.}
    \label{fig:GLEvsGFP_NR}
\end{figure}

We now introduce the exponential memory time in the equilibration process and solve the same system with a memory time $\tau_m=5$ fm in the generalized Langevin equation of Eq.~\eqref{eq:GLE_exp} in the non-relativistic case. The code now contains the ancillary (Ornstein-Uhlenbeck) process, which produces the correlated noise, as explained in Sec.~\ref{subsec:equilibration_memory}. The results are shown in Fig.~\ref{fig:GLEvsGFP_NR}.

In Fig.~\ref{fig:GLEvsGFP_NR} we use the same notation as in Fig.~\ref{fig:LEvsFP_NR} providing the momentum distribution as a function of elapsed time from the initial, Dirac delta condition. We can see that the evolution to equilibrium is slower than in the white noise case, since the dynamics are delayed due to the effect of memory. This effect has already been pointed out in Ref.~\cite{Ruggieri:2022kxv}. The generalized Fokker--Planck analytic results, shown in colored lines, perfectly match the numerical realization of the generalized Langevin equation.

The most remarkable effect is that the distribution seems to approach equilibrium around $t=8$ fm (cf. green line and gray dots), but does not stay nearby at later times. In fact, it bounces back to the red line (olive green dots) at $t=12$ fm, and then comes back to the equilibrium distribution at $t=40$ fm. The real-time solution shows that the distribution bounces back and forth around the equilibrium several times. The larger $\tau_m$ is, the more oscillations around it. 

The reason for this behavior is the presence of memory. Unlike the case without memory, the rate of change of momentum cannot be adjusted instantaneously. Instead, memory introduces a time delay in the system’s response, which effectively generates inertia-like behavior. As explained in the theory section, for an exponential memory kernel, the Langevin equation can be reduced to a form analogous to that of a damped harmonic oscillator. The particles relax toward the ``equilibrium'', but once they arrive there, the inertia drives the distribution past it, after which it returns with a delay effect. 

This can be more easily understood in the time evolution of the momentum component (with sign), instead of its modulus. In App.~\ref{app:GFP_1D} we provide the solution for the one-dimensional case in terms of the $z$ component of the momentum. We have enhanced the bouncing effect using a large value of $\tau_m=5$ fm. If this value is much smaller than the relaxation time $\tau_R=1/A$, then the dynamics are very close to the one in the memoryless case, without oscillatory behavior. Clearly, the distribution has approximately the correct shape, but it oscillates around the average momentum $\langle p_z \rangle=0$, since it carries an inertia due to memory time, which is corrected only after a time lapse.

\subsubsection{Relativistic evolution with colored noise~\label{sec:neq_rel_color}}

To conclude the nonequilibrium evolution, we provide the solution of the generalized Langevin equation for the relativistic case with particles of mass $m=1.27$ GeV and a bath temperature of $T=0.4$ GeV. In this case we do not compare to the solution of the generalized Fokker--Planck equation, since, as for the memoryless case, no analytical solution is available. While a numerical solution is certainly possible, it is beyond the scope of this work.

\begin{figure}[t!]
    \begin{center}
    \includegraphics[width=\linewidth]{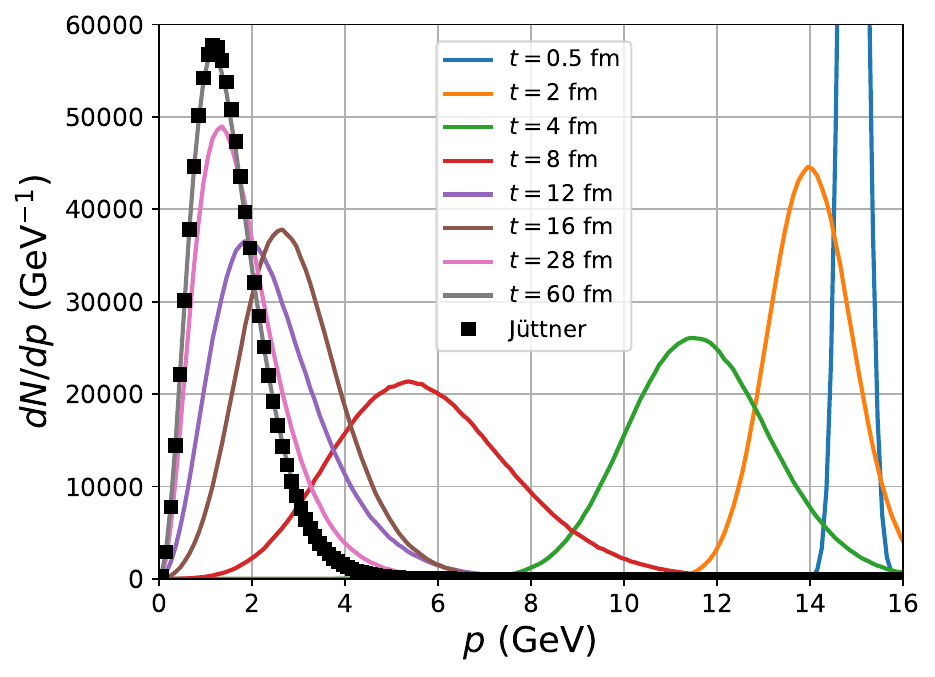}
    \end{center}
    \caption{
 Heavy-quark momentum distribution (\(\dd N/\dd p\)) evolution obtained from numerical generalized Langevin simulations in a relativistic system. Results are shown for a drag coefficient \(A=0.2\,\mathrm{fm}^{-1}\), temperature \(T=0.4\,\mathrm{GeV}\), mass \(m=1.27\,\mathrm{GeV}\), memory time \(\tau_m=5\,\mathrm{fm}\), an ensemble size of \(N=10^5\), and \(N_{\textrm{ev}}=50\) events.
    }\label{fig:GLE_N}
\end{figure}

In Fig.~\ref{fig:GLE_N} we present the distribution function evolution in time, showing the same equilibration process as the non-relativistic case. In this case, however, it is a non-Markovian dynamics, since the stochastic force exhibits temporal correlations over a finite memory time scale $\tau_m$, leading to a nonlocal evolution in time of the particle momentum, and, effectively, delaying the relaxation as compared to the Markovian case. 

The relativistic case also presents the oscillatory behavior of the solution around the ``equilibrium'' distribution. Here, it can be seen in the fact that the solution at $t=16$ fm has moved outwards from the equilibrium as compared to the solution at $t=12$ fm. Eventually, the solution tends to the true equilibrium solution, shown with square symbols in the plot. The result at $t=60$ fm is already close enough to it.

\subsection{Equilibrium: Correlation function and the Green--Kubo relation}

\subsubsection{Equilibrium in the white noise case}

In this section, we perform a series of calculations in the same box, but at thermal equilibrium at fixed temperature $T$. We start by applying the standard Langevin equation with uncorrelated noise. In the rest of this work, we exclusively work in the relativistic case with $m=1.27$ GeV and drag coefficient $A=0.2$ fm$^{-1}$.

\begin{figure}[t!]
    \begin{center}
     \includegraphics[width=\linewidth]{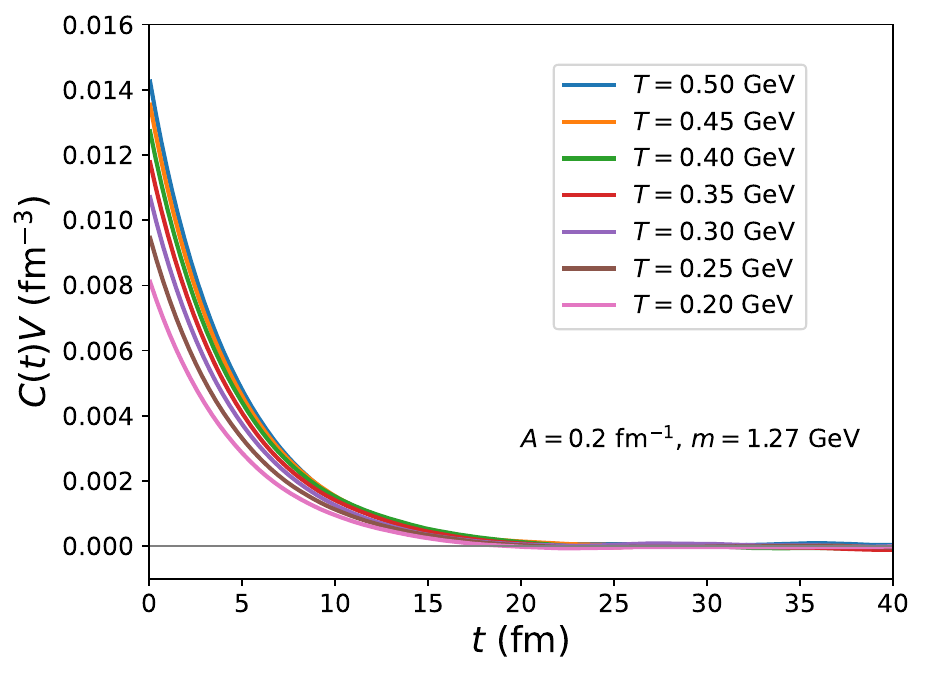}
    \includegraphics[width=\linewidth]{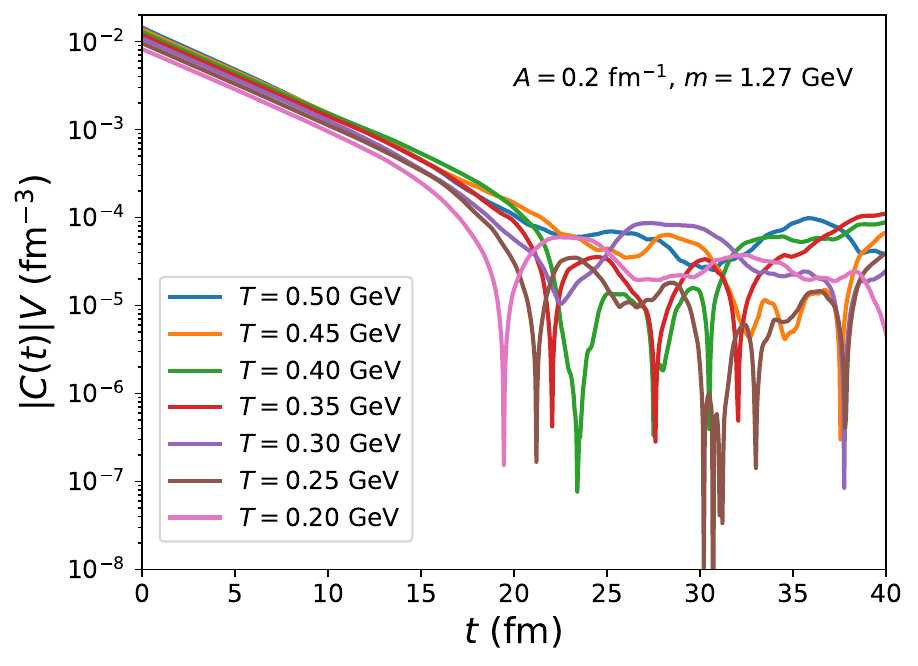}
    \end{center}
  \caption{Volume-scaled heavy-quark current--current correlation function \(C(t)\,V\) as a function of time in a static box at various temperatures, evaluated without memory effects.
 } \label{fig:corre_womem_rel_A}
 \end{figure}   

In Fig.~\ref{fig:corre_womem_rel_A} we present the current--current correlation function (multiplied by the volume of the system) as a function of time. Here $t$ denotes the time separation between two HQ currents in the correlator. In the upper panel, we show several correlation functions, for different temperatures, ranging from $T=0.2$ GeV to $T=0.5$ GeV. 

In the non-relativistic case, the correlation function is proven to be an exponential decay function, with inverse exponent $\tau_R=1/A$. However, as mentioned in Sec.~\ref{sec:GK_relativistic}, in the relativistic case, the exact form of the correlator is not known, but it is expected that its shape is still close to a decaying exponential, as can be seen in Fig.~\ref{fig:corre_womem_rel_A}. In the lower panel of this figure we plot the absolute value of the correlation function on a semi-logarithmic scale, where the exponential decay is evident for the first $\sim 10$ fm. After this time, the lack of statistics makes the correlation noisy, but it is already close to zero. In view of these results, we assume this exponential form in the relativistic case and compute the inverse slope $\tau$ from exponential fits to the first few fermi/$c$ of these $C(t)$.

 \begin{figure}[t!]
    \begin{center}
     \includegraphics[width=\linewidth]{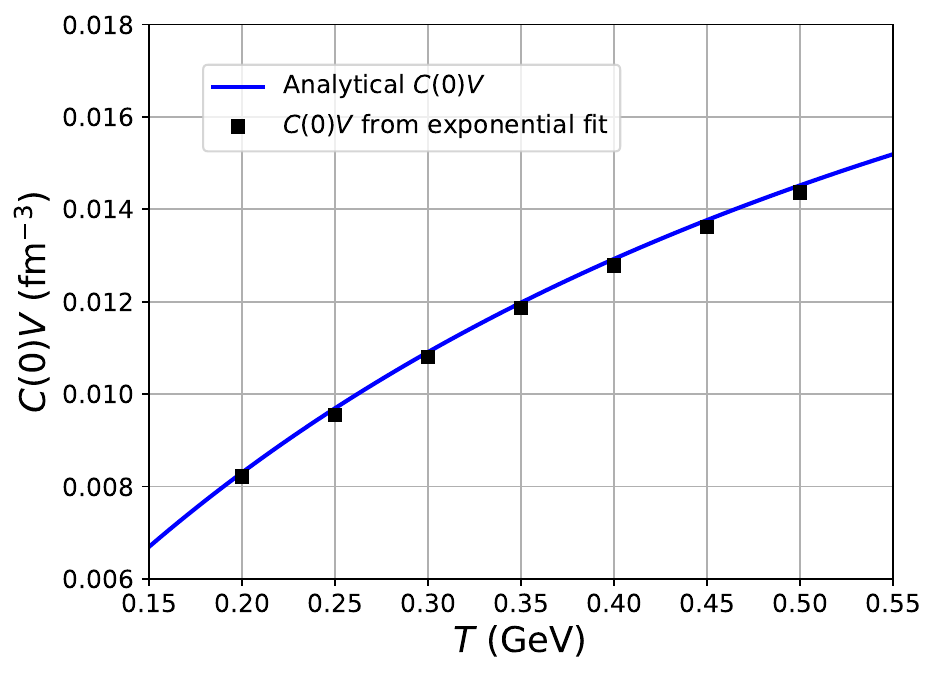}
     \includegraphics[width=\linewidth]{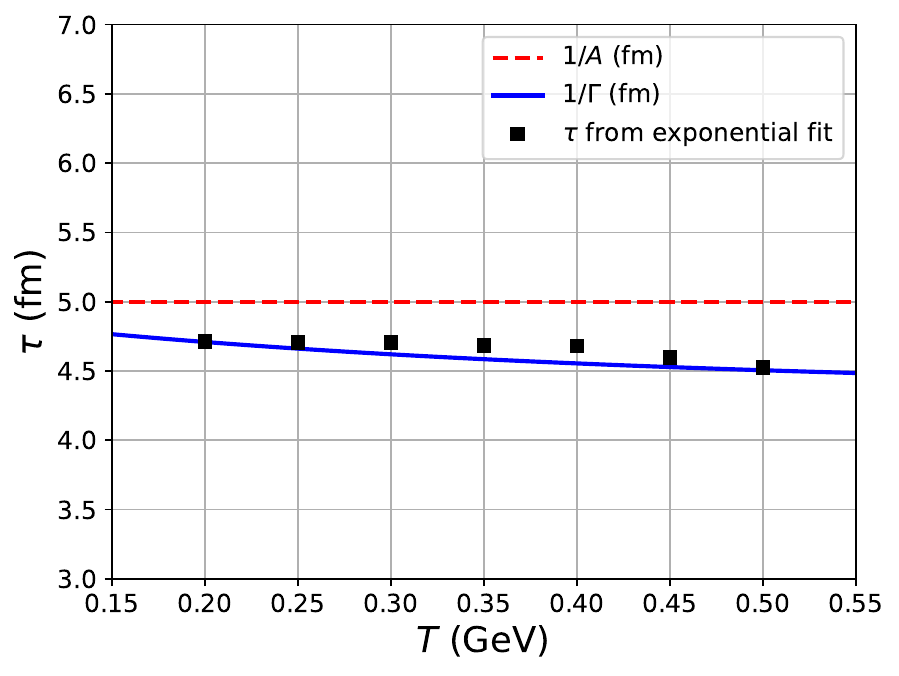}
    \end{center}
    \caption{$C(0)V$ (upper panel) and $\tau$ (lower panel) obtained from exponential fits to the current--current correlation functions for the uncorrelated noise case.} 
    \label{fig:tau_C0_womem_rel_A}
\end{figure}

After performing two-parameter fits in which $C(0)$ and $\tau$ are extracted for each temperature, in  Fig.~\ref{fig:tau_C0_womem_rel_A} we plot these two parameters as functions of the temperature. In the upper panel, we show $C(0)V$ for the seven temperatures analyzed together with the theoretical expectation given in Eq.~\eqref{eq:C0rel}. The agreement is excellent. In the lower panel, we present the inverse slope $\tau$ and compare it with the value of $\tau_R=1/A=5$ fm (constant for every temperature), and with the value of the inverse of the friction coefficient. In this case, since $\Gamma (E)$ is a function of the energy, we have used a thermally averaged one by combining the expression of $\Gamma(E)$ given in Table~\ref{tab:param} together with the average energy from Eq.~\eqref{eq:avE}. We observe that the fitted time $\tau$ is very close to this estimate $1/\Gamma(\langle E \rangle)$, as expected from the friction coefficient in the Langevin equation.

\begin{figure}[t!]
    \begin{center}
     \includegraphics[width=\linewidth]{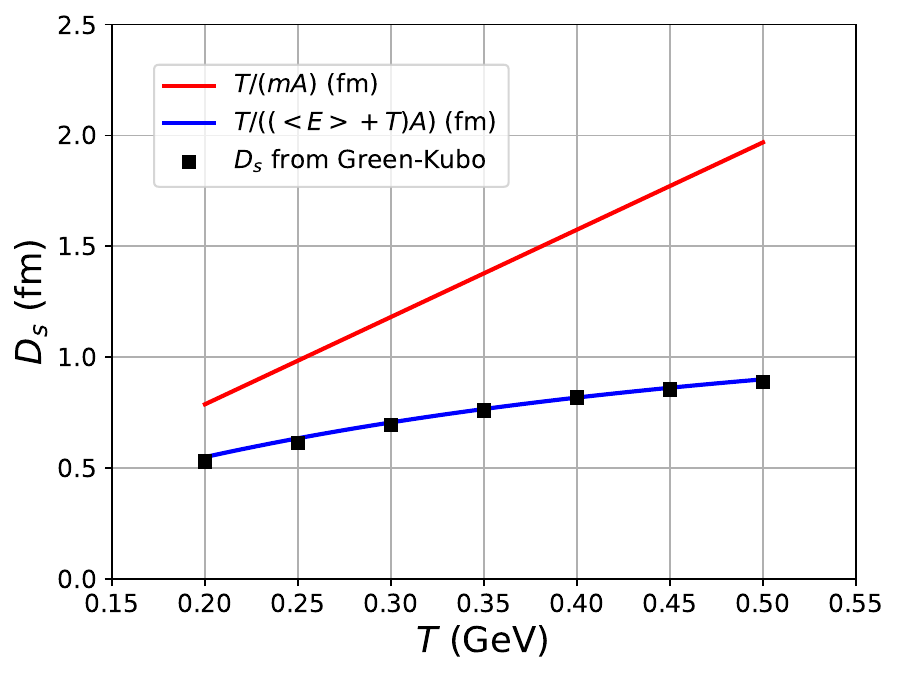}    \end{center}
    \caption{Spatial diffusion coefficient \(D_s\) at different temperatures, obtained from the Green--Kubo formula, Eq.~\eqref{eq:Dsrel}, for the uncorrelated noise case.} 
    \label{fig:Ds_womem_rel_A}
\end{figure}

Applying the Green--Kubo relation to the exponential form, we now compute the spatial diffusion coefficient $D_s$ from Eq.~\eqref{eq:Dsrel}. This expression is also valid in the non-relativistic limit (where $\tau$ would be exactly equal to $\tau_R=1/A$). The result is given in Fig.~\ref{fig:Ds_womem_rel_A} in square symbols for each of our seven temperatures.

In Fig.~\ref{fig:Ds_womem_rel_A} we also plot as a solid red line the theoretical expectation for the spatial diffusion coefficient using the non-relativistic expression in Eq.~\eqref{eq:DsA}. Of course, this equation is not valid in the relativistic case, and it does not compare well with our numerical solution. As mentioned previously, we know of no exact relation between $D_s$ and $A$ in the three-dimensional relativistic case in closed form. Alternatively, we can use the estimate for $D_s$ given by Eq.~\eqref{eq:einstein_relation}. We recall that this formula is just a heuristic estimate, extended from the NR version, using the appropriate Einstein relation for our case. This estimate is plotted as a blue solid line in Fig.~\ref{fig:Ds_womem_rel_A}. We observe excellent agreement between this expression for $D_s$ and the one calculated numerically from the Green--Kubo relation. This gives confidence that Eq.~\eqref{eq:einstein_relation} is close to the exact expression in the relativistic domain. We leave for the future the theoretical calculation of the exact expression from microphysics.

\subsubsection{Equilibrium in the colored noise case}

We present our results in the case with an exponential memory kernel, with a characteristic memory time $\tau_m$. In the non-relativistic case, we have checked that the momentum correlation functions describe either the overdamped or the underdamped solution as in Eqs.~\eqref{eq:CpmemoryOD} and~\eqref{eq:CpmemoryUD}, depending on the value of $\tau_m$. The current--current correlation function also inherits these forms. However, as in the uncorrelated-noise case, for the relativistic case one cannot find analytical expressions for $C(t)$. 

In this section, we work in the relativistic case assuming that the functional form of the non-relativistic correlation function is still valid in this case, as we will qualitatively observe. Nevertheless, we do not assign physical meaning to any of the fitted parameters, as we did in the memoryless case, but use the fitted form in the Green--Kubo formula to compute the diffusion coefficient $D_s$.

\begin{figure}[h!]
    \centering
\includegraphics[width=0.84\linewidth]{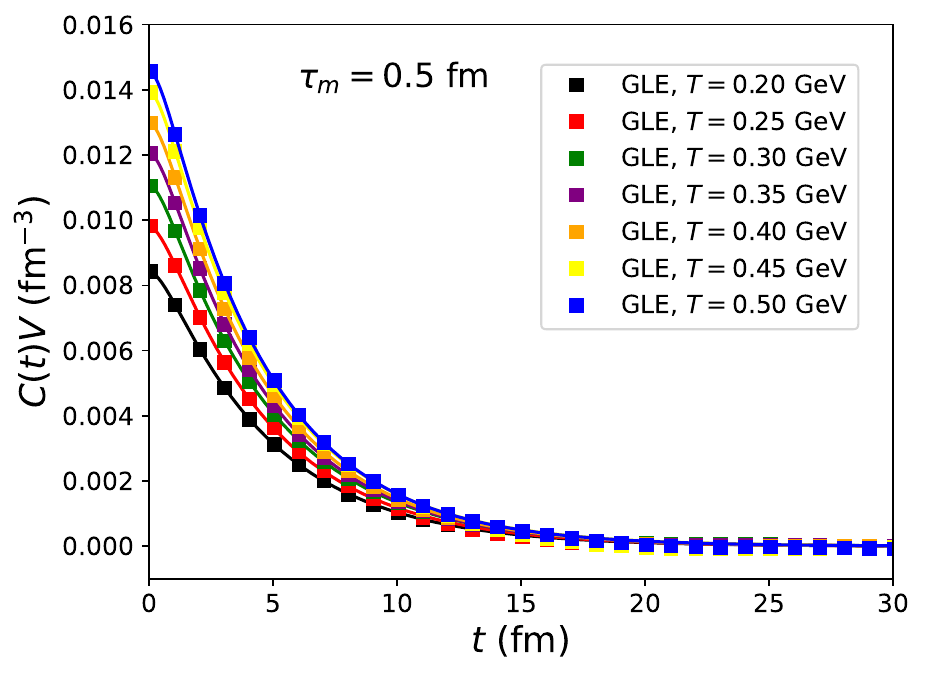}
\includegraphics[width=0.84\linewidth]{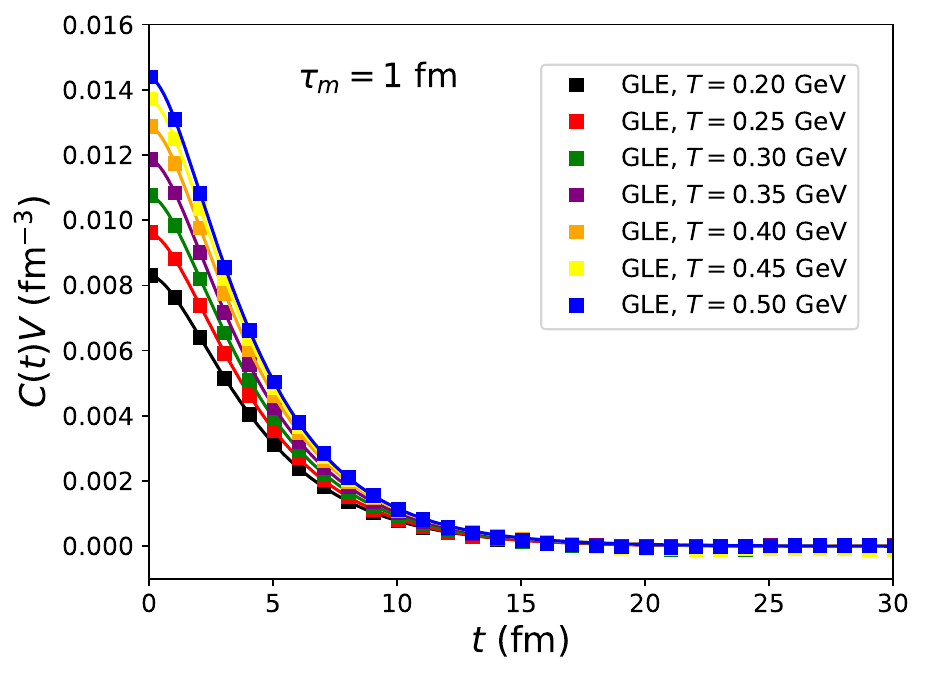} 
\includegraphics[width=0.84\linewidth]{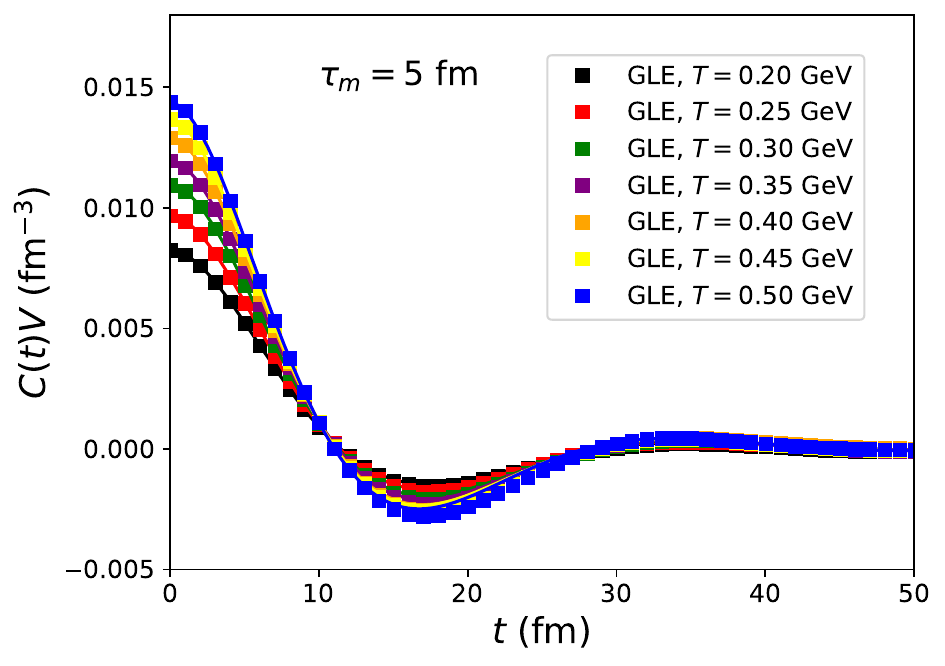}
 \includegraphics[width=0.84\linewidth]{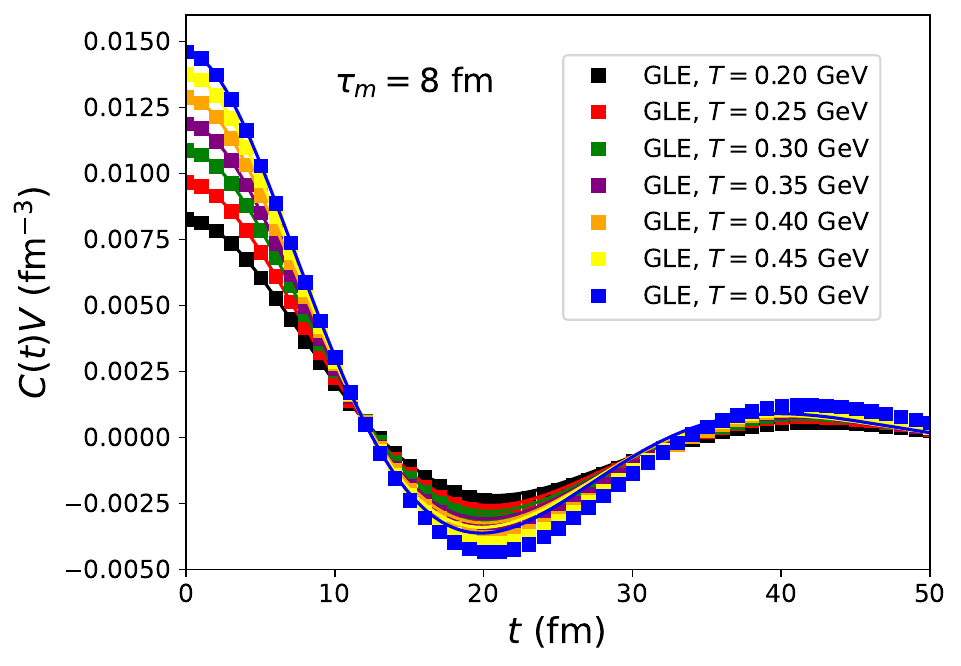}
   \caption{ Volume-scaled correlation function \(C(t)\,V\) as a function of time in a static box for different temperatures and memory times \(\tau_m\), evaluated for \(A=0.2\,\mathrm{fm}^{-1}\).
    \label{fig:corre_wm_rel_A}}
\end{figure}

A more agnostic approach could be to numerically integrate the correlation function without assuming any particular shape. However, the lack of statistics for large times in $C(t)$ introduces large systematic uncertainties. Nevertheless, we have done this exercise in the memoryless case in App.~\ref{app:intnum}. We observe a good agreement between the final results for the $D_s$ using the two methods.

We start by plotting the current correlation functions multiplied by the volume of the system $C(t)V$ in Fig.~\ref{fig:corre_wm_rel_A}. We use $m=1.27$ GeV, constant $A=0.2$ fm$^{-1}$, $V=(16 \textrm{ fm})^3$ and different temperatures ranging from $T=0.2$ GeV to $T=0.5$ GeV. We also scan the memory time $\tau_m$, starting with a small value of $\tau_m=0.5$ fm, where we almost recover the memoryless case, which corresponds to a  nontrivial check of the generalized Langevin equation with the ancillary process. We recall that this limit $\tau_m \rightarrow 0$ is ensured by the use of the Stratonovich-Fisk prescription, as dictated by the Wong-Zakai theorem~\cite{wong1969riemann}. 

Increasing the value of $\tau_m$ has a clear impact on the correlation function, making the system move from the overdamped case (for $\tau_m=0.5, 1$ fm) to the underdamped case (for $\tau_m=5, 8$ fm). In the latter case, we clearly observe oscillations around zero which are reminiscent of the nonequilibrium evolution seen in Sec.~\ref{sec:neq_rel_color}.

We qualitatively check that the values of $C(0)$ are essentially independent of $\tau_m$, but their temperature dependence is manifest.

In Fig.~\ref{fig:corre_wm_rel_A} we also plot the fitting curves (in solid colored lines) following the functional form provided by the exact non-relativistic results. Starting with the overdamped cases ($\tau_m=0.5,1$ fm), we apply the following fitting function
\be C(t)= C(0) \ e^{-t\alpha} \left[ \cosh (\beta t) + \gamma \ \sinh (\beta t) \right] \ , \label{eq:fit1}
\ee
with, in principle, four fitting parameters ($C(0),\alpha,\beta$ and $\gamma$). For the underdamped case ($\tau_m=5, 8$ fm)) we use
\be C(t)= C(0) e^{-t\alpha} \left[ \cos (\beta t) + \gamma \sin (\beta t) \right] \ , \label{eq:fit2}
\ee
with, again, in principle, four fitting parameters. While these forms are motivated by the non-relativistic expectations, we do not assume any relation of any of these parameters with the physical parameters used, as happens in the non-relativistic case. For example, it is tempting to associate $\alpha^{-1} =2\tau_m$, $\gamma=1/\sqrt{|\Delta|}$, and $\beta$ to $\sqrt{|1-4A\tau_m|}/(2\tau_m)=\alpha \gamma^{-1}$, as mandated in the non-relativistic case, but we do not do that in the relativistic case.

We find that the final results of the four-parameter fits are excellent, supporting the functional forms used for them. Moreover, even better results for $C(0)V$ can be obtained when this coefficient is not taken to be a fitting parameter, but simply fixed by the first point in the correlation function. Then, there are only three parameters in the fit: $\alpha, \beta, \gamma$. This method not only simplifies the fitting procedure by reducing the number of parameters, but is also preferred since $C(0)V$ is the coefficient affected by the least statistical uncertainty (it is the point computed with the largest number of pairs) and a global fit would treat all four parameters equally, losing the statistical significance of this parameter.

\begin{figure}[t!]
    \begin{center}
     \includegraphics[width=\linewidth]{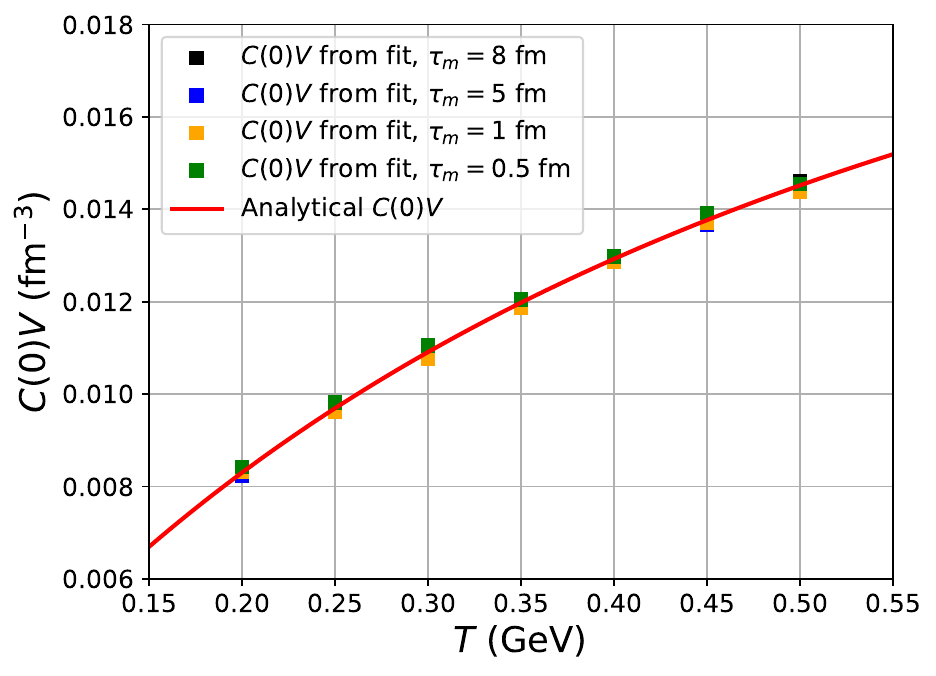}
    \end{center}
    \caption{Temperature dependence of the equal-time equilibrium correlator \(C(0)V\) for different memory times \(\tau_m\), evaluated at a fixed drag coefficient \(A=0.2\,\mathrm{fm}^{-1}\).} 
    \label{fig:C0V_rel_wmem}
\end{figure}

The result of the first value of the correlation function multiplied by the volume $C(0)V$ is presented in Fig.~\ref{fig:C0V_rel_wmem}. Clearly, $C(0)$ does not depend on $\tau_m$ (tiny variations are ascribed to numerical effects) and the temperature dependence agrees very well with the analytical prediction of Eq.~\eqref{eq:C0rel}. This follows from the fact that $C(0)$ is an equilibrium quantity, not related to the dynamical evolution of the system.

We finally consider the spatial diffusion coefficient obtained from the Green--Kubo relation. In the non-relativistic limit one would use Eq.~\eqref{eq:GK_memory} which already proved that $D_s$ does not depend on $\tau_m$, but only on $A$ and $T$. In the relativistic case, we also use the Green--Kubo technique applied to the analytical fitting functions~\eqref{eq:fit1} and~\eqref{eq:fit2}. In terms of the fitting parameters for each $T$ and $\tau_m$ we obtain
\be D_s = \frac{V}{3 q n_c} \int_0^\infty  \dd t \, C(t) = \frac{VC(0)}{3 q n_c} \frac{\alpha+ \beta \gamma}{\alpha^2 \mp \beta^2} \ , \ee 
where the minus (plus) sign in the denominator refers to the 
overdamped (underdamped) case.

\begin{figure}[t!]
    \begin{center}
     \includegraphics[width=\linewidth]{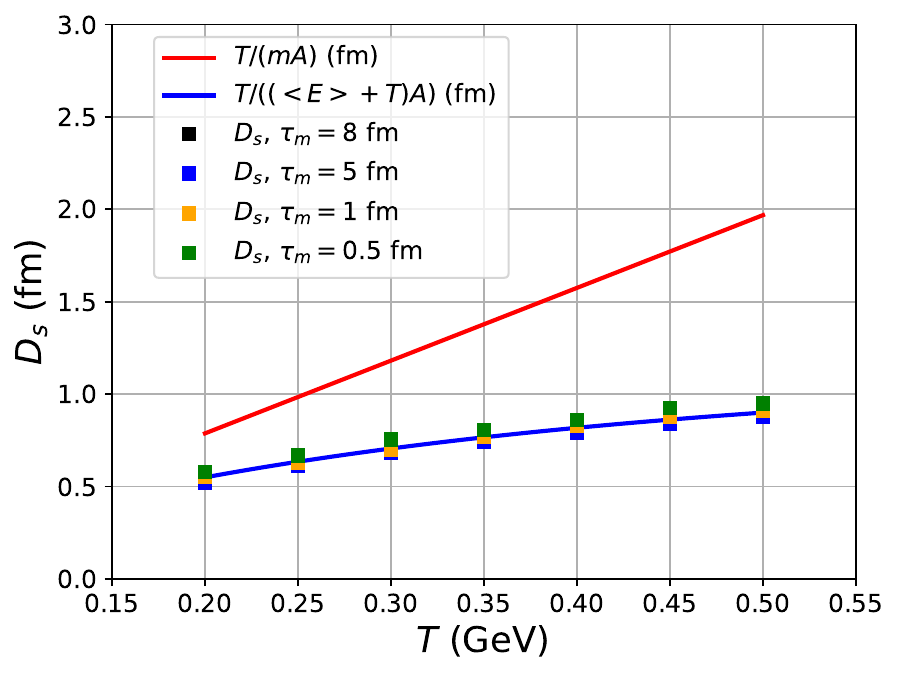}
    \end{center}
    \caption{Spatial diffusion coefficient \(D_s\) as a function of temperature for different memory times \(\tau_m\), evaluated at a fixed drag coefficient \(A=0.2\,\mathrm{fm}^{-1}\).} 
    \label{fig:Dsmemory}
\end{figure}

In Fig.~\ref{fig:Dsmemory} we present our result for $D_s$ as a function of $T$ and $\tau_m$. In this figure, we do not observe a significant dependence of the $D_s$ coefficient on $\tau_m$. The apparent tiny variations are compatible with numerical uncertainties. Moreover, we find that the obtained $D_s$ is compatible with the estimate given by Eq.~\eqref{eq:einstein_relation} and incompatible with the naive non-relativistic estimate $D_s=T/(mA)$. Therefore, we conclude that the Green--Kubo extraction of the $D_s$ coefficient with memory time is consistent with the values of the coefficient in the white noise case, and only dependent on $A$, $T$ and $m$ but not on $\tau_m$. 

At first sight, this claim might seem incompatible with the previous result in which the presence of a memory time delays the equilibration process. If the equilibration process gets slowed down, then one intuitively expects a smaller diffusion coefficient, when the memory time is larger.

In fact, this argument can be true in the \textit{transient} of the equilibration process, where an \textit{effective diffusion coefficient} would be different from the nominal $D_s$ calculated in the absence of any memory time. In the large-time limit of such a nonequilibrium evolution, we have proven that all solutions  eventually reach the same equilibrium distribution, independently of $\tau_m$, and independently of how many oscillations the distribution makes around the equilibrium. 

This ``large-time limit'' is inherent in the Green--Kubo formula, since it is rooted in hydrodynamics (valid for asymptotically-large time and length scales). This is manifest in the $+\infty$ upper limit in the integral of the Green--Kubo formula. Therefore, in spite of the transient dynamics of $C(t)$, and on how many oscillations it can make around zero, when integrating the correlation function until $t=+\infty$, the result leads to the same transport coefficient. In this sense, it is not surprising that $D_s$ should not depend on $\tau_m$ for a long-lived system in the thermodynamic limit.

What happens when the system is short lived, as in relativistic heavy-ion collisions? Then, the transient dominates and the relaxation of the system is truncated by the lifetime of the medium. Within the Green--Kubo approach, assuming that the system is not very far from equilibrium such that it keeps up with the equilibration at every time in the expansion, this can be captured by cutting off the upper limit of the integral, exactly as in the calculation of App.~\ref{app:intnum}. If one attempts to calculate $D_s$ using the Green--Kubo formula, but the system only lives, say until $t \simeq 5-10$ fm~\cite{Shen:2012vn}, the correlation function clearly does not capture the dynamics beyond this time interval, and the effective diffusion coefficient would be closer to $D_s(t^*=5-10 \textrm{ fm})$, different in magnitude than the hydrodynamic expectation $D_s(t^* = + \infty)$ as can be seen in Fig.~\ref{fig:Ds_numint_womem_rel_A}. For the particular cases in Fig.~\ref{fig:Ds_numint_womem_rel_A}, if the system's lifetime is around $t^*=5$ fm, the effective diffusion coefficient is 64\% of the hydrodynamic coefficient for the temperatures $T=0.2, 0.3, 0.4$ GeV; and 66\% for $T=0.5$ GeV. If the lifetime is doubled to $t^*=10$ fm, then more modes affecting the decay of the correlation function are active and the effective $D_s(t^*=10 \textrm{ fm})$ is now 87 \% of the hydrodynamic $D_s (t \rightarrow +\infty)$ for the temperatures $T=0.2,0.3,0.4$ GeV, and $88 \%$ for $T=0.5$ GeV.

The idea of an effective transport coefficient has been explored in Ref.~\cite{Rose:2020lfc} for the bulk viscosity and in Ref.~\cite{Pooja:2023gqt} for the heavy-flavor diffusion coefficient. A more detailed discussion of the effective diffusion coefficient in the context of realistic systems (like heavy-ion collisions) is postponed to a future publication,  since the values we have used here for $A,T,\tau_m$ might not necessarily be close to those in relativistic heavy-ion collisions. In any case, one always needs to assume that the system is not far from equilibrium and that the characterization in terms of a transport coefficient is possible. Otherwise, the linear response theory and the consistency of transport coefficients can break down.

\section{Summary and outlook~\label{sec:summary}} 

We have studied the late-time dynamics of heavy quarks in a thermal bath, both with and without memory effects, in the non-relativistic and relativistic regimes. The main focus is on the heavy-quark current-current correlation function and the thermalization process in the presence and absence of memory. The momentum evolution of heavy quarks is analyzed using Langevin dynamics, and the generalized Langevin equation is solved to incorporate time-correlated noise and thus account for memory effects. In this study, we consider an exponential decay of the memory kernel. For comparison with the Langevin approach, we also solve the Fokker--Planck equation to investigate heavy-quark momentum evolution with and without memory in the non-relativistic case.

We begin by studying Langevin dynamics in both the non-relativistic and relativistic regimes with constant drag coefficients to investigate heavy-quark momentum evolution. We perform the simulation at temperature $T = 0.1~\mathrm{GeV}$ with an initial momentum $p_0 = 5~\mathrm{GeV}$ for the non-relativistic case, whereas for the relativistic case we consider a temperature $T = 0.4~\mathrm{GeV}$ with an initial momentum $p_0 = 15~\mathrm{GeV}$.  The appropriate Einstein relation between the drag and diffusion coefficients is implemented, as discussed in Section~\ref{LV_wn}, and, as expected, the heavy quark approaches equilibrium at late times. We then extend the study to heavy-quark thermalization in the presence of memory effects by solving the generalized Langevin equation. 

In this framework, the time correlation of the thermal noise does not decay instantaneously but instead exhibits an exponential decay characterized by a timescale $\tau_m$, known as the memory time. Memory is introduced through an ancillary  stochastic process $h(t)$ that evolves independently of the heavy quarks, but in parallel with them, and is constructed such that its correlation at different times does not vanish. This process is then coupled to the Langevin equation to incorporate memory effects in the description of heavy-quark momentum evolution. 

We then solve the generalized Langevin equation, which incorporates memory effects, using the same temperature $T$, initial momentum $p_0$, and transport coefficients as in the case without memory. We observe that the evolution toward equilibrium in the presence of memory is slower than in the white-noise case, as the dynamics are delayed due to memory effects.

A notable feature is that the distribution initially approaches equilibrium, then moves away from it, and finally returns to equilibrium. This oscillatory behavior is seen in both the non-relativistic and relativistic cases. Moreover, the larger the memory time, the more pronounced is this oscillatory back-and-forth behavior, which can be understood as an effective inertia induced by memory effects. In the presence of memory, the Langevin equation becomes an integro-differential equation. For an exponential memory kernel, it can be reduced to two first-order differential equations, which combine into a second-order ordinary differential equation analogous to that of a damped harmonic oscillator. Since the momentum cannot adjust instantaneously, memory introduces a delayed response that effectively generates second-order, inertia-like dynamics. This oscillatory back-and-forth behavior is more clearly visible in the time evolution of the momentum itself rather than its magnitude. We present a snapshot for the one-dimensional case in App.~\ref{app:GFP_1D}, obtained using the Fokker--Planck equation.

We then compute the heavy-quark current-current correlation function in both the non-relativistic and relativistic cases without memory effects. In the non-relativistic case, the correlation function decays exponentially, with the relaxation time given by the inverse of the drag coefficient. Furthermore, the correlation function can be related to the transport coefficient through the Green--Kubo relation, which connects transport coefficients to the correlation functions of the corresponding conserved currents. 

We then extend the analysis to the relativistic case; however, the analytical expressions become more involved. We therefore employ the general Green--Kubo relation to compute $D_s$ from the current--current correlation function. In this case, we assume that the correlation function continues to exhibit an exponential decay when using white Gaussian noise in the Langevin equation, which can be evaluated at any temperature. Accordingly, we model the correlation function in an exponential form, where $\tau$ is treated as a fit parameter that characterizes the decay.

The correlation function, scaled by the system's volume, obtained from the analytical expression shows good agreement with that extracted from the exponential fit. We also compare the decay time $\tau$ obtained from the fit in the relativistic case with the inverse of the friction coefficient and the inverse of the drag coefficient. In the relativistic regime, these two are indeed different; however, the difference decreases as the temperature is lowered. The diffusion coefficient $D_s$ obtained from the Green--Kubo relation is compared with that estimated from the relativistic version of the Einstein relation, and we observe good agreement between the two.

We then compute the current--current correlation function with memory for different values of the memory time and at various temperatures. As memory time increases, the shape of the correlation function evolves from overdamped to underdamped dynamics, leading to negative correlations and oscillations around zero, reminiscent of nonequilibrium dynamics. We find that the equal-time correlation is independent of the memory time, up to numerical precision. Interestingly, in the presence of memory, the correlation function can be described by a sum of two exponentials, in contrast to the single-exponential behavior observed without memory. The diffusion coefficient $D_s$ with memory is then obtained using the Green--Kubo relation. A striking observation is that $D_s$ remains independent of memory time. Moreover, $D_s$ extracted from the Green--Kubo relation is in good agreement with that obtained from the Einstein relation in the relativistic case.

In conclusion, we have investigated heavy-quark dynamics in both Markovian and non-Markovian setups, considering both non-relativistic and relativistic cases. Our primary focus is on heavy-quark thermalization in the presence of memory, along with the analysis of the heavy-quark current--current correlation function to extract the diffusion coefficient $D_s$ using the Green--Kubo relation with memory effects. We plan to extend this approach in future works to explore heavy-quark phenomenology in a non-Markovian medium, including charmonium suppression in the quark--gluon plasma with memory effects.

\appendix

\section{Hydrodynamic fluctuations and the Green--Kubo formula}
\label{app:hydrofluc}

Transport coefficients can be formulated in terms of equilibrium correlation functions within the linear response theory~\cite{zwanzig1965time,landau1981physical,toda1992statistical,zubarev1996statistical}. For the heavy-flavor charge density $n_c(t,\bm{x})$ with chemical potential $\tilde{\mu}$, the associated macroscopic current $J^i (t,\bm{x})$ obeys, to leading order in gradients---after averaging over multiple realizations of the medium---Fick's diffusion law,
\begin{equation}
J^i(\bm{x},t) = D_s \nabla^i n_c (\bm{x},t)= \tilde{\chi} D_s \nabla^i \tilde{\mu} (\bm{x},t) \ , 
\label{eq:fick_law}
\end{equation}
where $\nabla^i=-\partial/\partial x^i$, $D_s$ is the spatial diffusion coefficient and $\tilde{\chi}$ is the susceptibility
\begin{equation}
\tilde{\chi} \equiv \left( \frac{\partial n_c}{\partial \tilde{\mu}} \right)_T \ ,
\label{eq:susceptibility}
\end{equation}
which, for a classical equilibrium distribution (Maxwell--Boltzmann or J\"uttner) is $\tilde{\chi}=qn_c/T$, since the only dependence on the chemical potential is through the fugacity factor $z=\exp(q\tilde{\mu}/T)$.

Using the theory of relativistic hydrodynamic fluctuations~\cite{fox1970contributions,Landau,Kapusta:2011gt, Kapusta:2012zb}, the local current exhibits fluctuations due to the dynamics of the microscopic degrees of freedom,
\begin{equation}
J^i(\bm{x},t)
= \tilde{\chi} D_s  \nabla^i \tilde{\mu}(\bm{x},t)
+ \xi^i(\bm{x},t) \ ,
\label{eq:local_current}
\end{equation}
with $\bm{\xi}(\bm{x},t)$ a Gaussian white noise term,
\begin{align}
\langle \xi^i(\bm{x},t) \rangle & = 0 \ , \\
\langle \xi^i(\bm{x},t) \xi^j(\bm{x}',t') \rangle
& =
2 T \tilde{\chi} D_s\,
\delta^{ij}
\delta^{(3)}(\bm{x}-\bm{x}')
\delta(t-t') \ .
\label{eq:noise_white}
\end{align}

More generally, the current can present a memory kernel,
\begin{equation}
J^i(\bm{x},t) =
\int \dd t' \, \dd^3x' \,
\Sigma^{ij}(\bm{x}-\bm{x}',t-t')
\,  \nabla_j \tilde{\mu}(\bm{x}',t')
+ \Xi^i(\bm{x},t) \ ,
\label{eq:nonlocal_current}
\end{equation}
where $\Sigma^{ij}$ is the retarded response kernel and $\Xi^i$ is a stochastic noise term, with 
\begin{align}
\langle \Xi^i(\bm{x},t) \rangle & = 0 \ , \\
\langle \Xi^i(\bm{x},t) \Xi^j(\bm{x}',t') \rangle
& =
T \ \Sigma^{*,ij} (\bm{x}-\bm{x}',t-t') \ .
\label{eq:noise_color}
\end{align}

The memoryless, Markovian limit corresponds to a local kernel~\cite{landau1981physical,Kapusta:2011gt},
\begin{equation}
\Sigma^{ij}(\bm{x}-\bm{x}',t-t')
= 2 \tilde{\chi} D_s \,
\delta^{ij}\delta^{(3)}(\bm{x}-\bm{x}')\delta(t-t') \ .
\label{eq:kernel_markov}
\end{equation}
This limit is difficult to reconcile with relativistic (retardation) effects, even in the case of non-relativistic particles. This is why non-local memory kernels have been considered in the context of heavy-ion collisions~\cite{Murase:2013tma,Kapusta:2014dja,Hammelmann:2018ath,Prakash:2026izm}.

We define the volume-averaged current,
\begin{equation}
j^i(t) \equiv \frac{1}{V}\int \dd^3x \, J^i(\bm{x},t) \ , 
\end{equation}
where the volume $V$ could be the entire volume of the system or a smaller part of it. Of course, in the latter case, fluctuations have a larger intensity.

Now, we work in an equilibrated system, i.e. in the absence of any external chemical potential gradient. However, the current is nonzero, since the stochastic term is still present. Locally, the current deviates from zero due to the random motion of the particles inside $V$, and any local charge imbalance is erased by the response of the system, controlled by the transport coefficient $D_s$.

This can be seen in the two-point correlation of the current,
\begin{equation}
C(t) \equiv \langle \bm{j} (t) \cdot \bm{j} (0) \rangle \ ,
\end{equation}
from which, in the memoryless case, one can trivially obtain the following relation by using Eqs.~\eqref{eq:local_current} and~\eqref{eq:noise_white},
\begin{equation}
\int_0^\infty \dd t \, C(t)=\int_0^{\infty} \dd t \,
\langle \bm{j}(t)\!\cdot\!\bm{j}(0) \rangle
= \frac{3 \tilde{\chi} D_s T}{V} \ .
\label{eq:current_integral}
\end{equation}

Then, we obtain for $D_s$,
\begin{equation}
D_s =
\frac{V}{3 T \tilde{\chi}}
\int_0^{\infty} \dd t \,
\langle \bm{j}(t)\!\cdot\!\bm{j}(0) \rangle \ , 
\label{eq:GK_Ds2}
\end{equation}
which is a Green--Kubo formula that relates a transport coefficient to the correlation function of the associated conserved current.

Fick's diffusion law relates the nonequilibrium response current with the initial gradient as an instantaneous process. Theoretically, the lack of retardation effects at the macroscopic level might be questionable. As mentioned, to account for a delayed reaction of the system, one can incorporate a memory kernel, e.g. in the form of an exponentially correlated noise, as done in Refs.~\cite{Murase:2013tma,Hammelmann:2018ath},
\begin{equation} \Sigma^{ij}(\bm{x}-\bm{x}',t-t')=\frac{D_s \tilde{\chi}}{\tau} \delta^{ij} \delta^{(3)} (\bm{x}-\bm{x}') \exp \left( - \frac{|t-t'|}{\tau} \right) \ ,
\end{equation}
where $\tau$ is (only in this appendix) a memory time between the gradient and the response current. In Ref.~\cite{Hammelmann:2018ath}, by one of us, we checked that this description is equivalent to having an exponentially decaying current with time $\tau$ together with a delta-correlated noise.

In this case, also in the absence of any external gradient, we identify the current with the stochastic term; and the current--current correlation function is  proportional to the noise two-point correlator~\eqref{eq:noise_color}, which now is an exponentially decaying function,
\begin{equation}
    C(t) = \frac{3 \tilde{\chi} D_s T}{V \tau} \exp \left( - \frac{|t|}{\tau} \right) \ . \label{eq:Ctcoloured}
\end{equation} 
In Ref.~\cite{Hammelmann:2018ath}, it was checked that this form was accurate enough to describe an interacting hadronic mixture, at least, in the dilute limit. Finally, it is possible to check that the correlation function in Eq.~\eqref{eq:Ctcoloured} also satisfies the Green--Kubo relation~\eqref{eq:GK_Ds2} for the coefficient $D_s$. 

Using the linear response theory~\cite{zwanzig1965time,landau1981physical}, the Green--Kubo formula is obtained rather generically without assuming any particular shape for the correlation function. On the other hand, the use of the theory of hydrodynamic fluctuations~\cite{Landau} assumes a particular form of the current--current correlator in equilibrium. The two formalisms are intimately related. The fact that the nonequilibrium response of the system to a local external gradient and the system's relaxation from a local fluctuation in equilibrium follow the same dynamics (and both are controlled by the same transport coefficient) is ensured by the microscopic universality and by the Onsager regression hypothesis~\cite{Onsager:1931jfa,Onsager:1931kxm}.

It is important to recall that the distinction between colored and white noise, in fact, depends on the time scale in which the fluctuations are measured. If the characteristic observing time is much longer than $\tau$ one can effectively treat the noise as white ($\tau \rightarrow 0$ limit).

\section{1D solution of the non-Markovian Fokker--Planck equation~\label{app:GFP_1D}}

To better understand the oscillatory behavior of the result of the generalized Fokker--Planck and Langevin equations in the presence of memory time in Sec.~\ref{sec:neq_rel_color},
we present in this appendix the numerical results in a more simplified, illustrative scenario. 

We consider a 1D nonequilibrium evolution along the $z$ direction and show $f(p_z,t)$ as a function of $p_z$, for both positive and negative regions of $p_z$, in  the non-relativistic limit, where an analytical expression can be obtained. 

We start with the analog of Eq.~\eqref{eq:GFP3D} in one dimension and write the solution of this GFP equation subject to the initial condition
$f(p_z,t=0)= \delta(p_z-p_0)$. It reads,
\be 
f(p_z,t) = \frac{1}{\sqrt{2 \pi  \sigma^2 (t)}} \exp \left\{ -\frac{[p_z-p_{0} \chi(t)]^2}{2 \sigma^2(t)} \right\} \ ,
\ee
with
\be \chi(t) = e^{-\frac{t}{2\tau_m}}  \left[ \cos \left( \frac{\sqrt{|\Delta|} t}{2\tau_m} \right) + \frac{1}{\sqrt{|\Delta|}} \sin \left( \frac{\sqrt{|\Delta|} t}{2\tau_m} \right) \right] \ , \ee
where $|\Delta| = 4A\tau_m-1$ and $\sigma^2(t)=mT[1-\chi^2(t)]$.

\begin{figure*}[!t]
    \centering

    \begin{subfigure}{0.48\textwidth}
        \centering
        \includegraphics[scale=0.63]{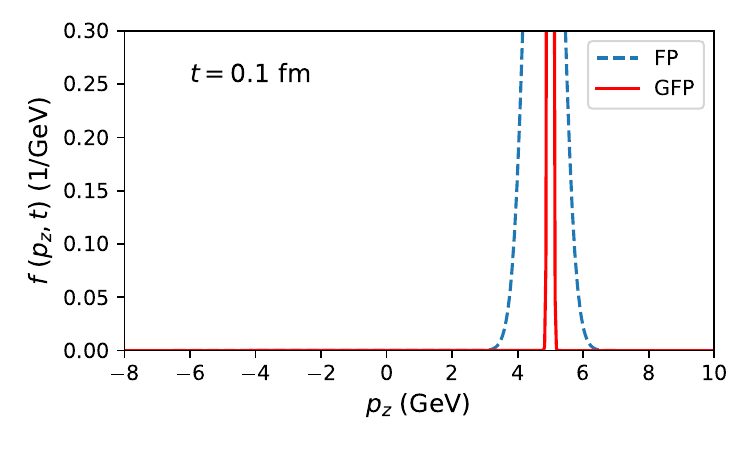}
    \end{subfigure}
    \begin{subfigure}{0.48\textwidth}
        \centering
        \includegraphics[scale=0.63]{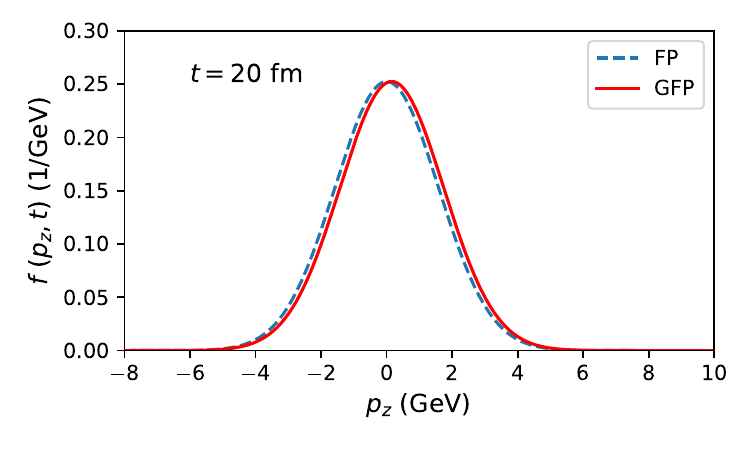}
    \end{subfigure}


      \begin{subfigure}{0.48\textwidth}
        \centering
        \includegraphics[scale=0.63]{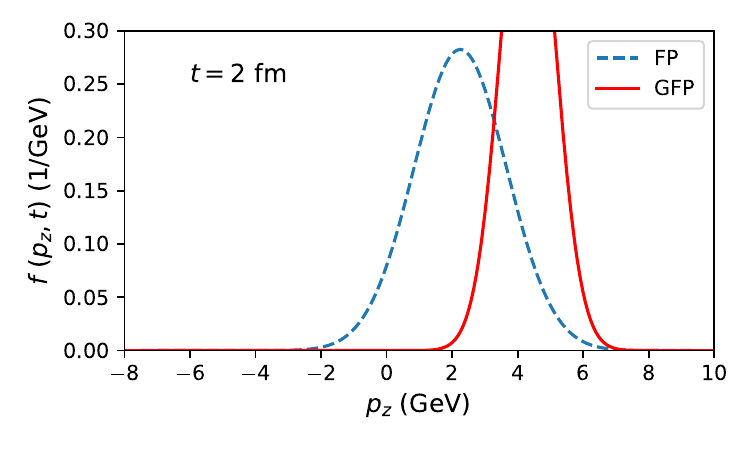}
    \end{subfigure}
    \begin{subfigure}{0.48\textwidth}
        \centering
        \includegraphics[scale=0.63]{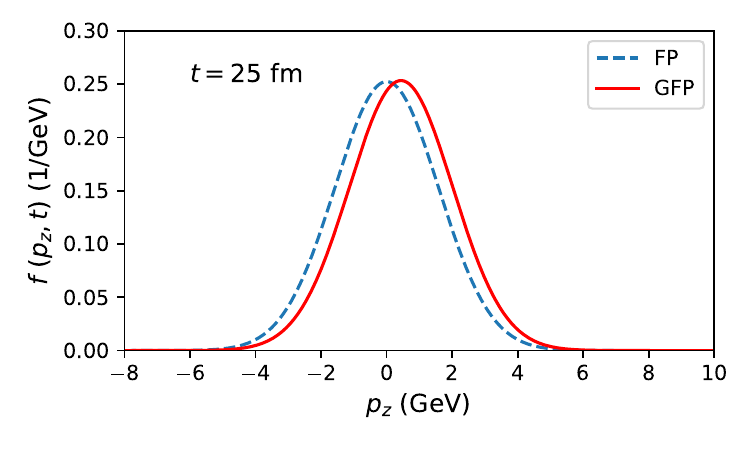}
    \end{subfigure}

      \begin{subfigure}{0.48\textwidth}
        \centering
        \includegraphics[scale=0.63]{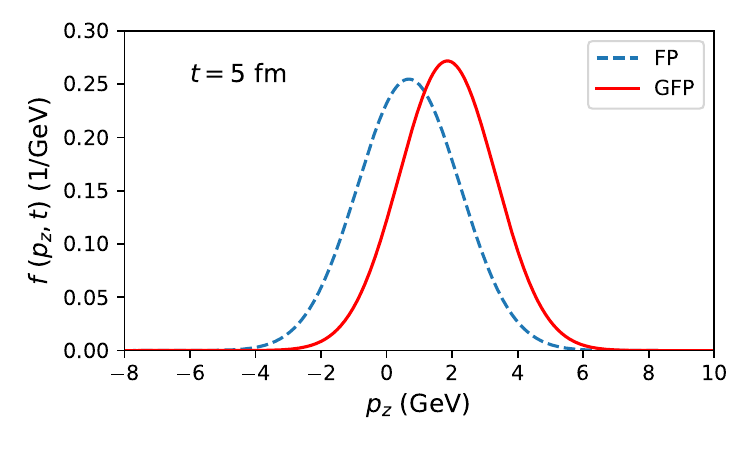}
    \end{subfigure}
    \begin{subfigure}{0.48\textwidth}
        \centering
        \includegraphics[scale=0.63]{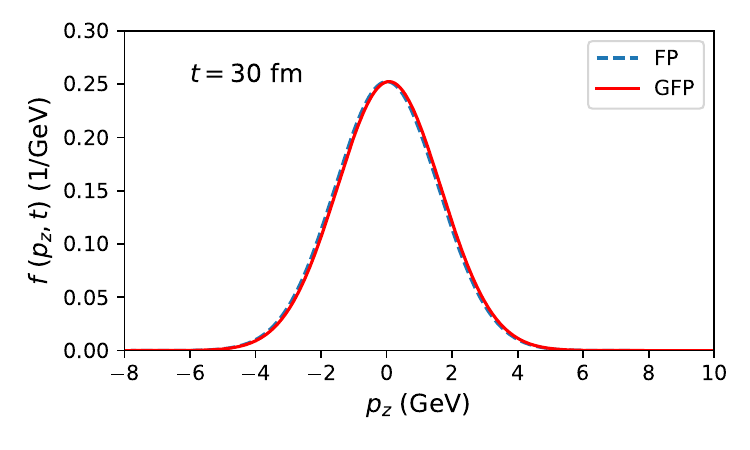}
    \end{subfigure}

      \begin{subfigure}{0.48\textwidth}
        \centering
        \includegraphics[scale=0.63]{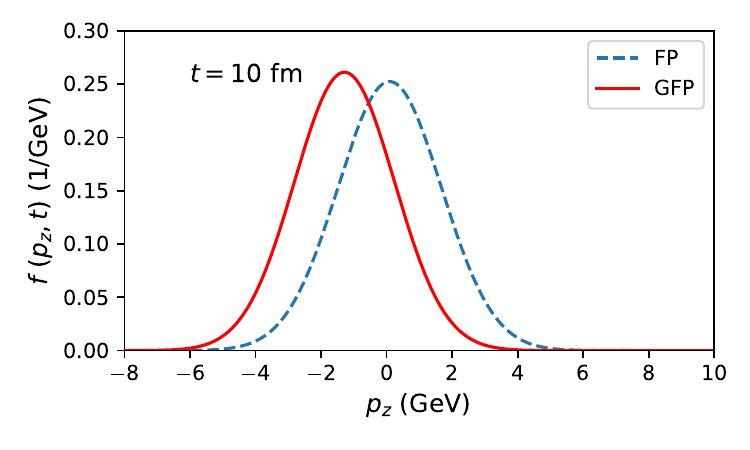}
    \end{subfigure}
    \begin{subfigure}{0.48\textwidth}
        \centering
        \includegraphics[scale=0.63]{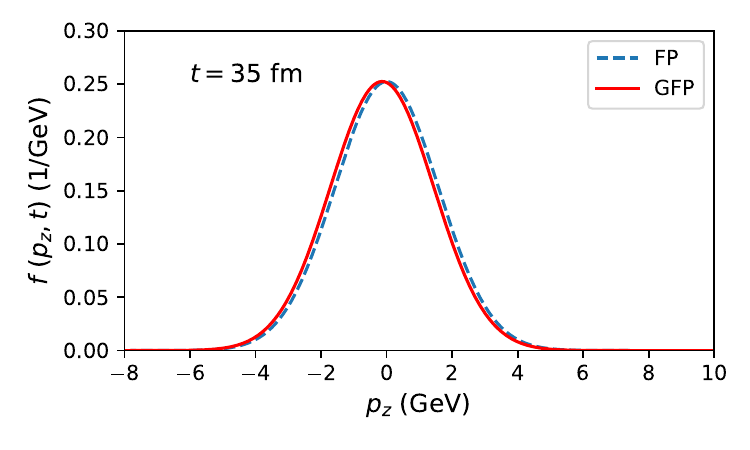}
    \end{subfigure}

      \begin{subfigure}{0.48\textwidth}
        \centering
        \includegraphics[scale=0.63]{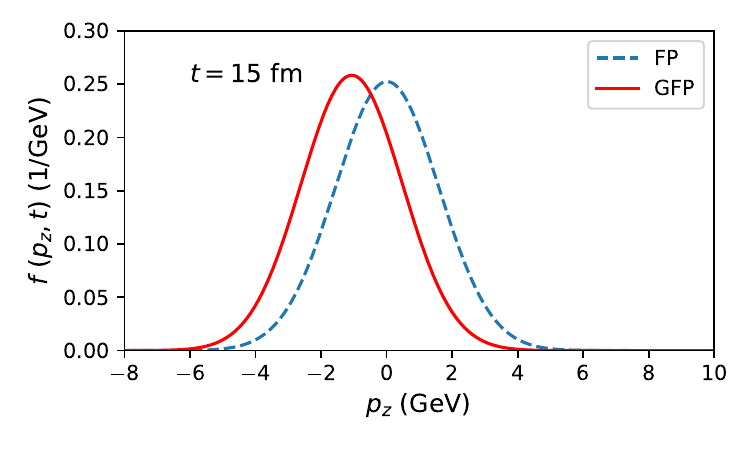}
    \end{subfigure}
    \begin{subfigure}{0.48\textwidth}
        \centering
        \includegraphics[scale=0.63]{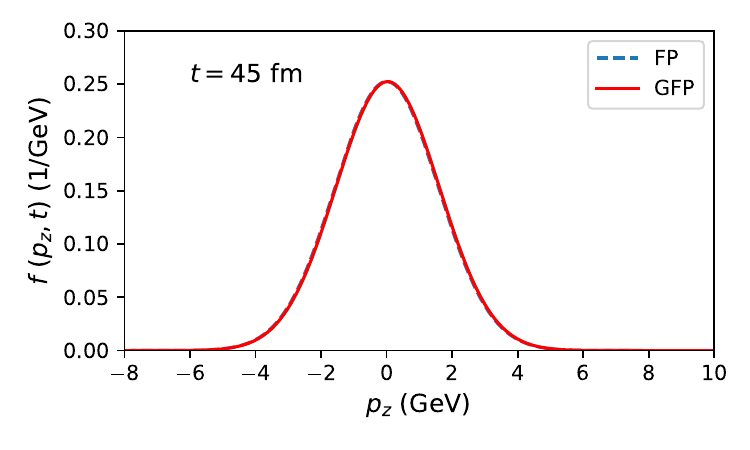}
    \end{subfigure}

    \caption{Heavy-quark momentum distribution \(f(p_z,t)\) as a function of \(p_z\) and time, obtained from solutions of the Fokker--Planck equation (blue dashed lines) and the generalized Fokker--Planck equation (red solid lines) in a non-relativistic system. Results are shown for a drag coefficient \(A=0.4\,\mathrm{fm}^{-1}\), temperature \(T=0.1\,\mathrm{GeV}\), mass \(m=25\,\mathrm{GeV}\), and memory time \(\tau_m=5\,\mathrm{fm}\).}
    \label{fig:1D_NR_memory_A}
\end{figure*}

We consider the following parameters: $A=0.4$ fm$^{-1}$, $m=25$ GeV, $T=0.1$ GeV, and $\tau_m=5$ fm. This solution is plotted as a red solid line in Fig.~\ref{fig:1D_NR_memory_A}, for several times from $t=0.1$ fm until $t=45$ fm. The solution of the (Markovian) FP equation is also plotted in blue dashed line for comparison. Time increases from top to bottom and from left to right panels.

The first feature to be noticed is the delay effect of the GFP solution, which is lagging behind the FP one, already at short times. The larger the memory time, the more delay the solution suffers. 

Second, we observe that around $t=5$ fm the FP distribution reaches the equilibrium solution and remains static. Since this is the true equilibrium distribution, it satisfies $\partial f/\partial t=0$. However, the GFP solution still needs several fermi to reach the true equilibrium. At $t=10$ fm we observe that the GFP solution has passed the equilibrium solution and continues to move toward negative values of the momentum. This is the inertia effect generated from the memory. At $t=15$ fm the GFP solution has turned back and approximates again the equilibrium distribution, which crosses again at $t=20$ fm. 
While at this time the GFP distribution function looks like the equilibrium one, it still has $\partial f/\partial t \neq 0$, and therefore is not yet the true equilibrium distribution. The oscillation around the equilibrium distribution continues several times more until the distribution approaches true equilibrium around $t=45$ fm, where now $\partial f/\partial t \simeq 0$.

\clearpage

\section{Numerical integration of correlation functions~\label{app:intnum}}

In this work, we assume certain analytical forms for the current--current correlation functions of relativistic systems, based on their non-relativistic exact expressions. That helps to exactly perform the Green--Kubo integration, but there is a limit in assessing the validity of these forms for all times, because of the lack of statistics at large times. In any case, the contribution of the large-time values of $C(t)$ to the transport coefficient is usually negligible.

Alternatively, one might rely on a more agnostic point of view and compute the associated diffusion coefficient from the Green--Kubo formula, using a numerical integration routine over the obtained $C(t)$ function,
\be D_s (t^*) = \frac{V}{3 q n_c} \int_0^{t^*} \dd t \, C(t)  \ . \ee 

In this equation, a sufficiently high upper cutoff $t^*$ is to be taken. Ideally, $D_s= \lim_{t^* \rightarrow \infty} D_s(t^*)$, but again, this formula assumes a good knowledge of the correlation function up to $t^*$, and for large times one simply runs out of statistics. However, this technique would still be useful when no functional form of $C(t)$ is known. 

In our Fig.~\ref{fig:corre_womem_rel_A} for the Markovian case, we obtain that for $t>15$ fm, the correlation functions are already close to zero, except for statistical fluctuations, which are difficult to reduce. We can monitor the result of $D_s(t^*)$ versus the upper cutoff $t^*$ and expect that the resulting integrated diffusion coefficient has reached a plateau. 

\begin{figure}[h!]
    \begin{center}
     \includegraphics[width=\linewidth]{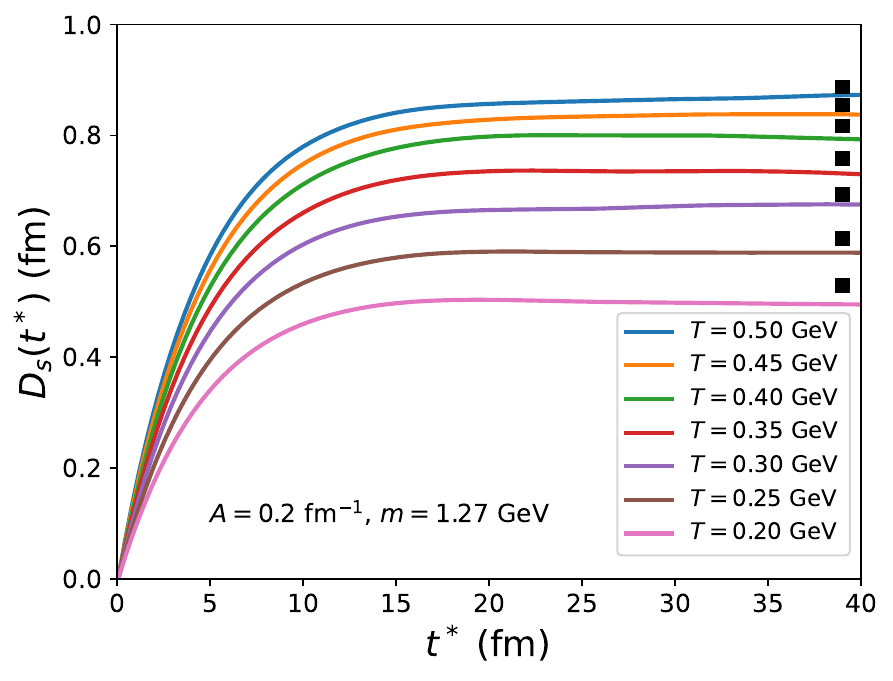}   
          \end{center}

    \caption{Integrated (or effective) diffusion coefficient \(D_s(t^*)\) for the white-noise case at different temperatures, with the upper cutoff \(t^*\) of the Green--Kubo formula for the correlation functions shown in Fig.~\ref{fig:corre_womem_rel_A}.     \label{fig:Ds_numint_womem_rel_A}}
\end{figure}

In Fig.~\ref{fig:Ds_numint_womem_rel_A} we show  $D_s(t^*)$ as a function of $t^*$ for the different temperatures considered in this work.

We observe a monotonic increase of the integrated coefficients until $t \simeq 15-20$ fm. Around $t^* \simeq 20$ fm, the curves reach respective plateaus, which can be associated with the asymptotic value of the transport coefficient. However, we note that due to the long tail of $C(t)$ which oscillates around zero, the asymptotic values in Fig.~\ref{fig:Ds_numint_womem_rel_A} fluctuate around the plateaus. This reflects the small ambiguity in choosing the optimal $t^*$.  

In symbols we also plot the results obtained in the main text assuming the exponential decay and performing the fit. The results are consistent for each temperature, with a systematic overestimation of the symbols with respect to their respective plateau. This is easily explained by the fact that while the integrated $D_s(t^*)$ accounts for both positive and negative values of $C(t)$ beyond $t=20-25$ fm, the calculation using the exponential shape only accounts for strictly positive values of $C(t)$, thus providing a systematically larger coefficient.

Finally, we notice that $D_s(t^*)$ can serve as the definition of an effective diffusion coefficient when the real system has a finite lifetime. If the system under study only lives for a certain time $t^*$, then a hydrodynamic response cannot be fully active and the transport coefficient cannot encode the response of the system beyond this $t^*$. The effective $D_s(t^*)$ would then be lower (in the memoryless case) than the one expected by the Green--Kubo with $t^* = + \infty$ in the full hydrodynamic regime. 

\begin{acknowledgments}

JMT-R acknowledges support from projects CEX2024-001451-M (Unidad de Excelencia ``Mar\'ia de Maeztu'') and PID2023-147112NB-C21, funded by the Spanish MCIN/ AEI/10.13039/501100011033/, and by Contract 2021 SGR 171 by the Generalitat de Catalunya, Grant No. 402942/2024-8 by the Brazilian CNPq (National
Council for Scientific and Technological Development).
SKD acknowledges useful discussions with Marco Ruggieri, Jai Prakash, and Pooja.
SKD acknowledges the support from Anusandhan National Research Foundation (ANRF), India, under grant No.:ANRF/ARG/2025/002424/PS.
The authors are grateful to the Mainz Institute for Theoretical Physics (MITP) of the Cluster of Excellence PRISMA+ (Project ID 390831469) for its hospitality and its partial support during the completion of this work. We also acknowledge the organizers of the 2026 MITP Workshop ``Exotic Quarkonia in Heavy-ion Collisions.''
\end{acknowledgments}

\bibliography{biblio_HQ_EM}

\end{document}